\documentclass[11pt]{article}
\usepackage{color}
\usepackage{amssymb}
\usepackage{amsfonts}
\usepackage{amsmath}
\usepackage{amsthm}
\usepackage[small,bf,figurewithin=section]{caption}
\usepackage{graphicx}
\usepackage{graphics}
\usepackage{epstopdf}
\usepackage{nicefrac}
\usepackage{multicol}
\usepackage{dsfont}
\usepackage{url}
\usepackage{capt-of}
\usepackage[english]{babel}

\title{Max-stable processes for modelling extremes observed in space and time}

\author{Richard A. Davis
\thanks{Department of Statistics, Columbia University, New York, United States, http://www.stat.columbia.edu/\textasciitilde rdavis, Email: rdavis@stat.columbia.edu} 
\and Claudia Kl\"uppelberg
\thanks{Center for Mathematical Sciences and Institute for Advanced Study, Technische Universit\"at M\"unchen, D-85748 Garching, Germany, http://www-m4.ma.tum.de/pers/cklu, Email: cklu@ma.tum.de}
\and Christina Steinkohl
\thanks{Center for Mathematical Sciences, Technische Universit\"at M\"unchen, D-85748 Garching, Germany, http://www-m4.ma.tum.de/pers/steinkohl, Email:steinkohl@ma.tum.de}}

\numberwithin{equation}{section}

\newtheorem{theorem}{Theorem}[section]
\newtheorem{lemma}[theorem]{Lemma}
\newtheorem{remark}[theorem]{Remark}

\newtheorem{proposition}[theorem]{Proposition}
\newtheorem{definition}[theorem]{Definition}
\newtheorem{corollary}[theorem]{Corollary}
\newtheorem{assumption}[theorem]{Assumption}
\newtheorem{fig}[theorem]{Figure}

\theoremstyle{definition}
\newtheorem{example}[theorem]{Example}

\newcommand{\bthe}{\begin{theorem}}
\newcommand{\ethe}{\end{theorem}}

\newcommand{\ben}{\begin{enumerate}}
\newcommand{\een}{\end{enumerate}}

\newcommand{\beq}{\begin{equation}}
\newcommand{\eeq}{\end{equation}}

\newcommand{\ble}{\begin{lemma}}
\newcommand{\ele}{\end{lemma}}

\newcommand{\bde}{\begin{definition}}
\newcommand{\ede}{\end{definition}}

\newcommand{\bco}{\begin{corollary}}
\newcommand{\eco}{\end{corollary}}

\newcommand{\bpr}{\begin{proposition}}
\newcommand{\epr}{\end{proposition}}

\newcommand{\bproof}{\begin{proof}}
\newcommand{\eproof}{\end{proof}}

\newcommand{\bexam}{\begin{example}}
\newcommand{\eexam}{\end{example}}

\newcommand{\bfi}{\begin{fig}}
\newcommand{\efi}{\end{fig}}

\newcommand{\btab}{\begin{tab}}
\newcommand{\etab}{\end{tab}}

\newcommand{\beao}{\begin{eqnarray*}}
\newcommand{\eeao}{\end{eqnarray*}\noindent}

\newcommand{\beam}{\begin{eqnarray}}
\newcommand{\eeam}{\end{eqnarray}\noindent}

\newcommand{\barr}{\begin{array}}
\newcommand{\earr}{\end{array}}

\newcommand{\bdis}{\begin{displaymath}}
\newcommand{\edis}{\end{displaymath}\noindent}
\newcommand{\bs}{\boldsymbol}

\newcommand{\bbn}{\mathbb{N}}

\newcommand{\bbr}{\mathbb{R}}

\newcommand{\la}{{\lambda}}

\newcommand{\si}{{\sigma}}

\begin{document}

\date{}
\maketitle

\begin{abstract}
Max-stable processes have proved to be useful for the statistical modelling of spatial extremes. Several representations of max-stable random fields have been proposed in the literature. For statistical inference it is often assumed that there is no temporal dependence, i.e., the observations at spatial locations are independent in time.  
We use two representations of stationary max-stable spatial random fields and extend the concepts to the space-time domain.  
In a first approach, we extend the idea of constructing max-stable random fields as limits of normalized and rescaled pointwise maxima of independent Gaussian random fields, which was introduced by Kabluchko, Schlather and de Haan \cite{Schlather2}, who construct max-stable random fields associated to a class of variograms. We use a similar approach based on a well-known result by H\"usler and Reiss \cite{Huesler} and apply specific spatio-temporal covariance models for the underlying Gaussian random field, which satisfy weak regularity assumptions. 
Furthermore, we extend Smith's storm profile model \cite{Smith} to a space-time setting and provide explicit expressions for the bivariate distribution functions. 
 
The tail dependence coefficient is an important measure of extremal dependence. We show how the spatio-temporal covariance function underlying the Gaussian random field can be interpreted in terms of the tail dependence coefficient. Within this context, we examine different concepts for constructing spatio-temporal covariance models and analyse several specific examples, including Gneiting's class of nonseparable stationary covariance functions \cite{Gneiting}. 

\end{abstract}
\vfill

\noindent
\begin{tabbing}
\
\> secondary: 62H11, 62M30  \,\,
\end{tabbing}

\vspace{1cm}


\vspace{0.5cm}

\section{Introduction}

The statistical modelling of extremes is an important topic in many applications. Environmental catastrophes like hurricanes, floods and earthquakes can cause substantial damage to structures like bridges, towers and buildings. Within an insurance context, those extremal events can result in large losses and it is essential to have an estimate of the risk of such catastrophes. Adequate stochastic models for the characterization and quantification of the behaviour of extremal events are needed, leading us to extreme value theory.  

Univariate extreme value theory is well developed and standard introductions can be found in many books, including for example Embrechts, Kl\"uppelberg and Mikosch~\cite{Embrechts}, Coles \cite{Colesbook} and Leadbetter, Lindgren and Rootz\'{e}n~\cite{Leadbetter}. The extension of the univariate to the multivariate extreme value distribution is considered for instance in de Haan and Resnick~\cite{deHaan4} and Beirlant et al. ~\cite{Beirlant}. In the simplest case, the process underlying the extremes is assumed to be independent or at least stationary and satisfying a mixing condition (see Leadbetter~\cite{Leadbetter1}). The development of nonstationary models for extremes is still in an evolutionary stage.
In the literature covariates are often introduced to overcome the nonstationarity of the underlying process. See for instance Davison and Smith~\cite{Davison} and Coles~\cite{Colesbook}, Chapter 6, for more details. 

Max-stable processes are a natural extension of multivariate extreme value distributions to infinite dimensions.
Detailed introductions and different spectral representations of stationary max-stable processes have been developed for example in Deheuvels \cite{Deheuvels}, de Haan \cite{deHaan}, de Haan and Pickands~\cite{deHaan3} and Schlather~\cite{Schlather1}. 
So far, max-stable processes have mostly been used for the statistical modelling of spatial data. Examples are given in Coles \cite{Coles2} and Coles and Tawn \cite{Coles3}, who model extremal rainfall fields using max-stable processes. Another application to rainfall data can be found in Padoan, Ribatet and Sisson \cite{Ribatet}, who also describe a practicable pairwise likelihood estimation procedure. 
An interesting application to wind gusts is shown in Coles and Walshaw~\cite{ColesWalshaw}, who use max-stable processes to model the angular dependence for wind speed directions.     

Different approaches for constructing max-stable processes have been introduced, and we mention two examples here, which we will extend to the space-time setting. In some cases the approaches can lead to the same finite-dimensional distribution functions.  

The idea of constructing max-stable random fields as limits of normalized and scaled pointwise maxima of Gaussian random fields was introduced in Kabluchko, Schlather and de Haan~\cite{Schlather2}, who construct max-stable random fields associated with a class of variograms. The limit field in this approach has the same finite dimensional distribution as the above described Brown-Resnick process (see Brown and Resnick~\cite{Brown} or Kabluchko et al.~\cite{Schlather2}).  

In an earlier paper Smith~\cite{Smith} introduced another max-stable process, which became known as the storm profile model. The different variables in the construction have an interpretation as components in a storm, including the shape and the intensity. The process is based on points from a Poisson random measure $\left\{(\xi_j,\bs{u}_j), j=1,\ldots\right\}$ together with a kernel function $f$, which in particular can be a centered Gaussian density. The max-stable process then arises from 
$\eta(\bs{y}) = \max_{j\geq 1} \xi_jf(\bs{y};\bs{u}_j)$. 

In real world applications, measurements are typically taken at various locations, sometimes on a grid, and at regularly spaced time intervals. For statistical inference, it is then often assumed, that the measurements are independent in time.  
We mention two approaches proposed in the literature concerning the analysis and quantification of the extremal behaviour of processes observed both in space and time. 
A first idea for modelling extremes in space and time can be found in Davis and Mikosch \cite{Davis}, who study the extremal properties of a moving average process, where the coefficients and the white-noise process depend on space and time.
Sang and Gelfand \cite{Sang} propose a hierarchical modelling procedure, where on a latent stage spatio-temporal dependence is included via the parameters of the generalized extreme value distribution. 
Extremes of space-time Gaussian processes have been studied in Kabluchko~\cite{Kabluchko1}. He analyses processes of the form 
$\sup_{t'\in [0,tn]}Z(s_n\bs{s},t')$ for some suitable chosen space-time Gaussian process and shows that the finite dimensional distributions of a properly scaled version converge to those of a Brown-Resnick process.

For the extension of Kabluchko et al.'s approach to a space-time model we need an underlying spatio-temporal correlation model for the Gaussian random field, which satisfies a certain regularity condition at 0. This condition, taken from a well-known result by H\"usler and Reiss~\cite{Huesler}, assumes that a rescaled version of the correlation function $\rho(\bs{h},u)$ for the space-lag $\bs{h}$ and the temporal lag $u$ satisfies $\log n(1-\rho(s_n\bs{h},t_n u)) \to \delta(\bs{h},u)$, as $n\to \infty$ for the scaling sequences $s_n$ and $t_n$. 
We establish an explicit connection between the limit $\delta(\bs{h},u)$ and the tail dependence coefficient for two locations at two time points.

Recently, the development of covariance models in space and time has received much attention and there is now a large literature available for the construction of a wide-range of spatio-temporal covariance functions. Examples can be found in Gneiting~\cite{Gneiting}, Ma~\cite{Ma1,Ma2}, 
Cressie and Huang~\cite{Cressie} and Schlather~\cite{Schlather3}. 
Within this context, we generalize an assumption on correlation functions for the analysis of extremes from stationary Gaussian processes (see for instance Leadbetter et al.{}\cite{Leadbetter}, Chapter 12), which is a sufficient condition for the limit assumption described above. 
We show how Gneiting's class of nonseparable and isotropic covariance functions \cite{Gneiting} fits into this framework. 

In addition, we examine spatial anisotropic correlation functions, which allow for directional dependence in the spatial components. 
Perhaps the easiest way to introduce anisotropy in a model is to use geometric anisotropy. For a detailed introduction, we refer to Wackernagel \cite{Wackernagel}, Chapter 9. Using this concept in the underlying correlation function, we can model anisotropy in the corresponding max-stable random field. 
Furthermore, we revisit a more elaborate way of constructing anisotropic correlation models based on Bernstein functions introduced by Porcu, Gregori and Mateu \cite{Porcu}, called the Bernstein class. 

Our paper is organized as follows. Two approaches for the construction of space-time max-stable random fields are described in Section 2. 
The max-stable random field, which is based on the maximum of rescaled and transformed replications of Gaussian space-time random fields is developed in Section \ref{HueslerReiss}, while Smith's storm profile model is extended to a space-time setting in Section \ref{Smithext}. In Section \ref{Pickands}, we show how Pickands dependence function and the tail dependence coefficient relate to the correlation model used in the underlying Gaussian random field. Further correlation models are discussed in Section \ref{Posscov} and simulations based on a set of different parameters are visualized. Section \ref{aniso} analyzes anisotropic correlation functions, where one can see directional movements in the storm profile model, which are not possible in the isotropic case.

\label{Introduction}

\section{Extension of extreme spatial fields to the space-time setting}
Max-stable processes form the natural extension of multivariate extreme value distributions to infinite dimensions. In the literature, different approaches for establishing space-time max-stable processes have been considered. In this section, we discuss two approaches for constructing such random fields and extend the concepts to a space-time setting. In Section~\ref{HueslerReiss} we describe the construction introduced in Kabluchko, Schlather and de Haan~\cite{Schlather2} based on a limit of pointwise maxima from an array of independent Gaussian random fields. 
Furthermore, we extend the approach introduced in de Haan~\cite{deHaan} and interpreted by Smith~\cite{Smith} as the storm profile model to a space-time setting in Section~\ref{Smithext}. 

\subsection{Max-stable random fields based on spatio-temporal correlation functions}\label{HueslerReiss}
Before presenting the construction of a max-stable Gaussian random field in space and time, we begin with the definition of the Brown-Resick space-time process with Fr\'echet marginals (see Brown and Resnick~\cite{Brown} or Schalther~\cite{Schlather1}). 
Let $\left\{\xi_j,j\geq 1\right\}$ denote points of a Poisson random measure on $[0,\infty)$ with intensity measure $\xi^{-2}d\xi$ and let $Y_j(\bs{s},t),j=1,2,\ldots$ be independent replications of some space-time random field $\left\{Y(\bs{s},t),(\bs{s},t)\in\bbr^d\times [0, \infty)\right\}$ with $\mathbb{E}(Y(\bs{s},t)) < \infty,$ and $Y(\bs{s},t)\geq 0 \ a.s.$, which are also independent of $\xi_j$. The random field, defined by 
\begin{equation}
\eta(\bs{s},t) = \bigvee\limits_{j=1}^\infty \xi_jY_j(\bs{s},t), \quad (\bs{s},t) \in \bbr^d\times [0,\infty)
\label{BrownResnick}
\end{equation}
is a max-stable random field with Fr\'echet marginals and often refered to as the Brown-Resnick process (see Kabluchko et al.~\cite{Schlather2})
The finite-dimensional distributions can be calculated using a point process argument as done in de Haan~\cite{deHaan}. For example, if $(\bs{s}_1,t_1),\ldots,(\bs{s}_K,t_K)$ are distinct space-time locations (duplicates in the space or the time components are allowed), then
\begin{align}
&P(\eta(\bs{s}_1,t_1)\leq y_1,\ldots, \eta(\bs{s}_K,t_K)\leq y_K) = P\left(\xi_j\bigvee\limits_{k=1}^K \frac{Y_j(\bs{s}_k,t_k)}{y_k}\leq 1, \forall j=1,2,\ldots\right)\nonumber \\
={}& P\left( N(A) = 0\right) =
\exp\left\{-\mathbb{E}\left(\bigvee\limits_{k=1}^K\frac{Y(\bs{s}_k,t_k)}{y_k}\right)\right\},
\label{BRfinite}
\end{align} 
where $A= \left\{(u,v);uv\leq 1\right\}$ and $N$ is the Poisson random measure with points at $$\left\{\left(\xi_j,\bigvee\limits_{j=1}^K\frac{Y_j(\bs{s}_k,t_k)}{y_k}\right)\right\}.$$
As we will see below, the Brown-Resnick process will be the limit of a sequence of pointwise maxima of independent Gaussian space-time processes. 
In the following, let $\left\{Z(\bs{s},t)\right\}$ denote a space-time Gaussian process on $\bbr^d\times[0,\infty)$ with covariance function given by
$$\tilde{C}(\bs{s}_1,t_1;\bs{s}_2,t_2) = \mathbb{C}ov\left(Z(\bs{s}_1,t_1),Z(\bs{s}_2,t_2)\right),$$ 
for two locations $\bs{s}_1,\bs{s}_2 \in \bbr^d$ and time points $t_1,t_2 \in [0,\infty)$.
We assume stationarity in space and time, so that we can write 
$$\tilde{C}(\bs{s}_1,t_1;\bs{s}_2,t_2) = C(\bs{s}_1-\bs{s}_2,t_1-t_2) = C(\bs{h},u),$$
where $\bs{h}=\bs{s}_1-\bs{s}_2$ and $u=t_1-t_2$.
Furthermore, let $\rho(\bs{h},u) = C(\bs{h},u)/C(\bs{0},0)$ denote the corresponding correlation function. 
We will assume smoothness conditions on $\rho(\cdot,\cdot)$ near $(\bs{0},0)$. This assumption is natural in the context of spatio-temporal random fields, since it basically relates to the smoothness of the underlying random field in space and time. 
\begin{assumption}\label{asscov}
There exists two nonnegative sequences of constants $s_n\to 0, \ t_n \to 0$ as  $n\to \infty$ and a nonnegative function $\delta$ such that 
\begin{equation*}
\log n(1- \rho(s_n (\bs{s}_1-\bs{s}_2),t_n (t_1-t_2))) \to \delta(\bs{s}_1-\bs{s}_2,t_1-t_2) \in (0,\infty),\quad  n\to \infty,
\end{equation*}
for all $(\bs{s}_1,t_1)\neq (\bs{s}_2,t_2),\ (\bs{s}_1,t_1),(\bs{s}_2,t_2) \in \bbr^d\times [0,\infty)$. 
\end{assumption}
Examples of such correlation functions are given in Section \ref{Posscov}.
If $(\bs{s}_1,t_1)=(\bs{s}_2,t_2)$, it follows that the correlation function equals one, $\rho(\bs{0},0) =1$, which directly leads to $\delta(\bs{0},0)=0$. 
The following theorem regarding limits of finite-dimensional distributions stems from Theorem 1 in H\"usler and Reiss \cite{Huesler}. In the following let $C(\bbr^d\times [0,\infty))$ denote the space of continuous functions on $\bbr^d\times [0,\infty)$, where convergence is defined as uniform convergence on compact subsets $K$ of $\bbr^d\times [0,\infty)$. 

\begin{theorem}\label{bivariateHuesler}
Let $Z_j(\bs{s},t),\ j=1,2,\ldots$ be independent replications from a stationary space-time Gaussian random field with mean $0$, variance $1$ and correlation model $\rho$ satisfiying Assumption~\ref{asscov} with limit function $\delta$. Assume there exists a metric $D$ on $\bbr^d\times[0,\infty)$ such that
\begin{equation}
\delta(\bs{s}_1-\bs{s}_2,t_1-t_2) \leq  (D((\bs{s}_1,t_1),(\bs{s}_2,t_2)))^2
\label{metric}
\end{equation}
and set
\begin{equation}
\eta_n(\bs{s},t) = \frac{1}{n} \bigvee\limits_{j=1}^n-\frac{1}{\log\left(\Phi(Z_j(s_n\bs{s},t_n t))\right)}, \quad  (\bs{s},t)\in \bbr^d\times [0,\infty).
\label{etanfrechet}
\end{equation}
Then,
\begin{equation}\label{limitHuesler}
\eta_n(\bs{s},t) \stackrel{\mathcal{L}}{\longrightarrow} \eta(\bs{s},t),
\end{equation}
where $\stackrel{\mathcal{L}}{\longrightarrow}$ denotes weak convergence in $C(\bbr^d\times [0,\infty))$ and $\left\{\eta(\bs{s},t),(\bs{s},t)\in\bbr^d\times[0,\infty)\right\}$ is a space-time max-stable process. 
The bivariate distribution functions for $\eta(\bs{s},t)$ have an explicit form given by
\begin{equation}\label{bivhuesler}
F(y_1,y_2) =\exp\left\{-\frac{1}{y_1}\Phi\left(\frac{\log\frac{y_2}{y_1}}{2\sqrt{\delta(\bs{h},u)}} + \sqrt{\delta(\bs{h},u)}\right) - \frac{1}{y_2}\Phi\left(\frac{\log\frac{y_1}{y_2}}{2\sqrt{\delta(\bs{h},u)}}+ \sqrt{\delta(\bs{h},u)}\right)\right\}.
\end{equation}
\end{theorem} 

\begin{remark}
Condition \eqref{asscov} is sufficient to prove tightness of the sequence $(\eta_n)_{n\in\bbn}$ in $C(\bbr^d\times [0,\infty))$.
As shown in the proof of Theorem 17 in Kabluchko et al.~\cite{Schlather2} the limit process $\eta$ turns out to be a Brown-Resnick process with $Y$ in \eqref{BrownResnick} given by 
$$\exp\left\{W(\bs{s},t)-\delta(\bs{s},t)\right\},\ (\bs{s},t)\in\bbr^d\times[0,\infty),$$ 
where $\left\{W(\bs{s},t), \ (\bs{s},t)\in\bbr^d\times [0,\infty)\right\}$ is a Gaussian random field with mean $0$ and covariance function
\begin{equation}\label{brkab}
\mathbb{C}ov(W(\bs{s}_1,t_1),W(\bs{s}_2,t_2)) = \delta(\bs{s}_1,t_1)+\delta(\bs{s}_2,t_2)-\delta(\bs{s}_1-\bs{s}_2,t_1-t_2).
\end{equation}
In particular, $\delta$ is a variogram leading to a valid covariance function in \eqref{brkab}. 
\end{remark}
\begin{proof}
Although this proof is similar to the one given in Kabluchko et al.~\cite{Schlather2}, we provide a sketch of the argument for completeness. We start with the bivariate distributions. 
From classical extreme value theory (see for example Embrechts, Kl\"uppelberg and Mikosch \cite{Embrechts}, Example 3.3.29), we have for
\begin{equation}\label{bn}
b_n = \sqrt{2\log n} -\frac{\log\log n+\log (4\pi)}{2\sqrt{2\log n}}
\end{equation} that
$$\lim_{n\to \infty}\Phi^n\left(b_n+\frac{\log(y)}{b_n}\right) =e^{-1/y}.$$
By using the standard arguments as in Embrechts et al. \cite{Embrechts}, it follows that
$$\Phi^{-1}\left(e^{-1/ny}\right)\sim \frac{\log y_1}{b_n} + b_n.$$
By applying this relation and Theorem 1 in H\"usler and Reiss \cite{Huesler} to the random variables $\eta_n(\bs{s}_1,t_1)$ and $\eta_n(\bs{s}_2,t_2)$ for fixed $(\bs{s}_1,t_1),(\bs{s}_2,t_2)\in \bbr^d\times[0,\infty)$, we obtain
\begin{align*}
&P(\eta_n(\bs{s}_1,t_1) \leq y_1, \eta_n(\bs{s}_2,t_2) \leq y_2)\\ 
={} & P\left(\bigvee_{j=1}^n -\frac{1}{\log(Z_j(s_n\bs{s}_1,t_n t_1))}\le ny_1, \bigvee_{j=1}^n  -\frac{1}{\log(Z_j(\bs{s}_2,t_n t_2))} \le ny_2\right)\\
={} & P\left(\bigvee_{j=1}^n Z_j(s_n\bs{s}_1,t_n t_1)\le \Phi^{-1}\left(e^{-1/(ny_1)}\right), \bigvee_{j=1}^n Z_j(\bs{s}_2,t_n t_2) \le \Phi^{-1}\left(e^{-1/(ny_2)}\right)\right)\\
\sim{} & P^n\left(Z_1(s_n\bs{s}_1,t_n t_1) \leq \frac{\log(y_1)}{b_n} + b_n, Z_1(s_n\bs{s}_2,t_n t_2) \leq \frac{\log(y_2)}{b_n} + b_n\right) \\
\sim{} & \exp\left\{-\frac{1}{y_1} - \frac{1}{y_2} + nP\left(Z_1(s_n\bs{s}_1,t_n t_1)>\frac{\log(y_1)}{b_n} + b_n , Z_1(s_n\bs{s}_2,t_n t_2)>\frac{\log(y_2)}{b_n} + b_n\right)\right\} \\ 
\end{align*}
The vector $(Z_1(s_n\bs{s}_1,t_n t_1),Z_2(s_n\bs{s}_2,t_n t_2))$ is bivariate normally distributed with mean $\bs{0}$ and covariance matrix given by $\rho(s_n(\bs{s}_1-\bs{s}_2),t_n(t_1-t_2))$. 
Using the properties of the conditional normal distribution and Assumption~\ref{asscov}, it can be shown that the last expression converges to \eqref{bivhuesler}.
Similarly to the procedure above, the finite-dimensional limit distributions of beyond second order can be calculated by using Theorem 2 in H\"usler and Reiss \cite{Huesler}.

It remains to show that the sequence $(\eta_n)$ is tight in $C(\bbr^d\times[0,\infty))$. Following Kabluchko et al.~\cite{Schlather2}, the main step of the proof is to show that the conditional family of processes $\left\{Y_n^{\omega}(\bs{s},t),\ (\bs{s},t)\in\bbr^d\times[0,\infty)\right\}$, given by
$$Y_n^{\omega}(\bs{s},t) = (b_n(Z(s_n\bs{s},t_nt)-b_n) - \omega)\mid (b_n(Z(\bs{0},0)-b_n = \omega), \quad \omega\in[-c,c], \ n\in \bbn$$ 
is tight in $C(K)$, where $K$ is any compact subset of $\bbr^d\times[0,\infty)$ and $b_n$ is defined in \eqref{bn}. This is achieved by calculating an upper bound for the variance of the distance between the process at two spatio-temporal locations, which in our case is given by Assumption \eqref{metric}. That is, for large $n$, 
\begin{align*}
\mathbb{V}ar(Y_n^{\omega}(\bs{s}_1,t_1) - Y_n^{\omega}(\bs{s}_2,t_2)) &\leq 2 b_n^2(1-\rho(s_n(\bs{s}_1-\bs{s}_2),t_n(t_1-t_2)))\\
&\leq 2B \delta(\bs{s}_1-\bs{s}_2,t_1-t_2) \leq 2BD((\bs{s}_1,t_1),(\bs{s}_1,t_2))^2,
\end{align*}    
where $B>0$ is some constant. The rest of the proof follows analogously to the proof in \cite{Schlather2}.  
\end{proof}

\begin{remark}
Kabluchko~\cite{Kabluchko1} studies the limit behavior of rescaled space-time processes of the form $$\sup_{t'\in[0,tn]}Z(s_n\bs{s},t'),$$
and shows that a rescaled version converges in the sense of finite-dimensional distributions to a space-time Brown-Resnick process. The assumptions on the covariance function in the underlying Gaussian space-time random field are similar to those we use in Section \ref{Posscov}. The approach differs from ours in the sense that we analyse the pointwise maxima of independent replications of space-time random fields, rather than the supremum over time of a single random field. 
\end{remark}

\begin{remark}
In applications, the marginal distributions are often fitted by a generalized extreme value distribution and are then transformed to standard Fr\'echet. 
Sometimes it may be useful to think about other marginal distributions, such as the Gumbel or Weibull. 
In order to use Gumbel marginals, we need 
\begin{equation}
\eta_n(\bs{s},t) = \bigvee\limits_{j=1}^n -\log\left(-\log\left(\Phi\left(Z_j(s_n\bs{s},t_n t)\right)\right)\right) - \log(n),
\label{etangumbel}
\end{equation}
and obtain the bivariate distribution function in \eqref{bivhuesler} with $1/y_1$ and $1/y_2$ replaced by $e^{-y_1}$ and $e^{-y_2}$. 
If we want to use Weibull marginals in our model, we obtain 
\begin{equation}
\eta_n(\bs{s},t) = n\bigvee\limits_{j=1}^n \log\left(\Phi\left(Z_j(s_n\bs{s},t_n t)\right)\right),
\label{etanweibull}
\end{equation}
leading to the same bivariate distribution function as in \eqref{bivhuesler}, but with $1/y_1$ and $1/y_2$ replaced by $y_1$ and $y_2$, respectively. 
\end{remark}

\subsection{Extension of the storm profile model}\label{Smithext}
In this section, we extend the following max-stable process, first introduced in de Haan \cite{deHaan}, to the space-time setting. 
The process was interpreted by Smith \cite{Smith} as a model for storms, where each component can be interpreted as elements of a storm, like intensity or center.   
In later papers, including for instance Schlather and Tawn \cite{SchlatherTawn} this process is called the storm profile model. 
We extend the concept to a space-time setting, where extremes are observed at certain locations through time. 
For simplicity of presentation we assume without loss of generality that $\bbr^2$ is the space domain.  
Assume, that we have a domain for point processes of storm centres $Z\subset \bbr^2$ and a time domain $X\subset [0,\infty)$, for which the storm is strongest at its centres.  
Further, let $\left\{(\xi_j,\bs{z}_j,x_j),j\geq 1\right\}$ denote the points of a Poisson random measure on $(0,\infty)\times Z\times X$ with intensity measure $\xi^{-2}d\xi\times \lambda_2(d\bs{z})\times \lambda_1(dx)$, where $\lambda_d$ denotes Lebesgue measure on $\bbr^d$ for $d=1,2$. Each $\xi_j$ represents the intensity of storm $j$.  
Moreover, let $f(\bs{s},t;\bs{z},x)$ for $(\bs{s},t)\in \bbr^d \times [0,\infty)$ and $(\bs{z},x)\in Z\times X$ be a non-negative function with
$$\int\limits_{Z\times X}f(\bs{s},t;\bs{z},x)\lambda_2(d\bs{z})\lambda_1(dx) = 1, \quad (\bs{s},t) \in \bbr^2 \times [0,\infty).$$
The function $f$ represents the shape of the storm. 
Define 
\begin{equation}
\eta(\bs{s},t) = \bigvee\limits_{j\geq 1}\left\{\xi_j f(\bs{z}_j,x_j;\bs{s},t)\right\}, \quad (\bs{s},t)\in \bbr^2 \times [0,\infty).
\label{smith}
\end{equation}
The product $\xi_jf(\bs{z}_j,x_j;\bs{s},t)$ can be interpreted as the wind speed at location $\bs{s}$ and time point $t$ from storm $j$ with intensity $\xi_j$, spatial location of the center $\bs{z}_j$ and maximum wind speed at time $x_j$ at the centre. 
The finite dimensional distribution function of $(\eta(\bs{s}_1,t_1),\ldots,\eta(\bs{s}_K,t_K))$, defined for fixed $(\bs{s}_1,t_1),\ldots,(\bs{s}_K,t_K) \in \bbr^d \times [0,\infty)$ and $y_1,\ldots,y_K \in \bbr$, is given through the spectral representation calculated in de Haan \cite{deHaan} as
\begin{eqnarray}
F(y_1,\ldots,y_K) &=& \exp\left\{-\int\limits_{Z\times X}\bigvee\limits_{k=1}^K\frac{f(\bs{z},x;\bs{s}_k,t_k)}{y_k} \, \lambda_2(d\bs{z})\lambda_1(dx)\right\}. 
\label{smithK}
\end{eqnarray}

To connect the storm model with the Brown-Resnick process that arises in Theorem \ref{bivariateHuesler}, we assume a trivariate Gaussian density for the function $f$ with mean $(\bs{z},x)$ and covariance matrix $\tilde{\Sigma}$, i.e. 
$$f(\bs{s},t;\bs{z},x) = f_0(\bs{z}-\bs{s},x-t),$$
where $f_0$ is a Gaussian density with mean $\bs{0}$ and covariance matrix $\tilde{\Sigma}$. 
We assume that the spatial dependence is modelled through the matrix $\Sigma$ and the temporal dependence is given through $\sigma_3^2$, leading to the covariance matrix
\beam\label{Smithcovmodel}
 \tilde{\Sigma} = 
\begin{pmatrix}
\Sigma & \bs{0}\\ \bs{0}& \sigma_3^2 	
\end{pmatrix} = 
\begin{pmatrix}
\sigma_1^2 & \sigma_{12} & 0\\ \sigma_{12} & \sigma_2^2 & 0 \\ 0 & 0& \sigma_3^2 	
\end{pmatrix}.
\eeam

In the following theorem, we calculate a closed form of the bivariate distribution function resulting from the setting defined above. The derivation of the bivariate distribution function in a purely spatial setting can be found in Padoan, Ribatet and Sisson \cite{Ribatet} and we stick closely to their notation. The idea of the proof is widely known and for completeness, details are given in Appendix \ref{Appendix1}. 
\begin{theorem}\label{bivariateSmith}
With the setting defined above, the max-stable space-time random field 
\begin{equation}\label{Smith2}
\eta(\bs{s},t) = \bigvee\limits_{j\geq 1}\left\{\xi_j f_0(\bs{z}_j-\bs{s};x_j-t)\right\}, \quad (\bs{s},t)\in S\times T,
\end{equation}
has the bivariate distribution function given by
\begin{align}\label{spacetimeds}
& F(y_1,y_2) = P(\eta(\bs{s}_1,t_1)\leq y_1,\eta(\bs{s}_2,t_2)\leq y_2) \nonumber \\ &= \exp\left\{-\frac{1}{y_1}\Phi\left(\frac{2\sigma_3^2\log(y_2/y_1) + \sigma_3^2a(\bs{h})^2 + u^2}{2\sigma_3\sqrt{\sigma_3^2a(\bs{h})^2+u^2}}\right) - \frac{1}{y_2}\Phi\left(\frac{2\sigma_3^2\log(y_1/y_2) + \sigma_3^2a(\bs{h})^2 + u^2}{2\sigma_3\sqrt{\sigma_3^2a(\bs{h})^2+u^2}}\right)\right\},
\end{align}
where $\bs{h} = \bs{s}_1-\bs{s}_2$ is the space lag, $u=t_1-t_2$ is the time lag and $a(\bs{h}) = (\bs{h}^T\Sigma^{-1}\bs{h})^{1/2}$.
\end{theorem}

Note, that if the time lag $u$ equals zero, the formula reduces to 
$$F(y_1,y_2) = \exp\left\{-\frac{1}{y_1} \Phi\left(\frac{a(\bs{h})}{2} + \frac{\log(y_2/y_1)}{a(\bs{h})}\right)- \frac{1}{y_2}\Phi\left(\frac{a(\bs{h})}{2} + \frac{\log(y_1/y_2)}{a(\bs{h})}\right)\right\},$$ 
which is the bivariate distribution of a Gaussian max-stable random field in space as calculated in Padoan, Ribatet and Sisson \cite{Ribatet}.
If the space lag $\bs{h}$ is zero, the bivariate distribution function is given by
$$F(y_1,y_2) = \exp\left\{-\frac{1}{y_1} \Phi\left(\frac{1}{u}\left(\log(y_2/y_1) + \frac{u^2}{2\sigma_3^2}\right)\right) - \frac{1}{y_2} \Phi\left(\frac{1}{u}\left(\log(y_1/y_2) + \frac{u^2}{2\sigma_3^2}\right)\right)\right\}.$$

By comparing the bivariate distributions from the Smith model with those of the approach discussed in \eqref{bivhuesler} in Section~\ref{HueslerReiss}, we recognize that the functions are the same, if 
\begin{equation}\label{SmithHuesler}
\delta(\bs{h},u) = \frac{1}{4}a(\bs{h})^2 + \frac{1}{\sigma_3^2}u^2.
\end{equation}
In Section~\ref{Posscov}, where we study a more detailed representation of the function $\delta$, we come back to this point.

\section{Pickands dependence function and tail dependence coefficient} \label{Pickands}
The Pickands dependence function (Pickands \cite{Pickands}) is one measure of tail dependence and is related to the so-called exponent measure. A general introduction to exponent measures and Pickands dependence function can be found in Beirlant et al.~\cite{Beirlant}. In particular, the joint distribution of the max-stable random field can be expressed with the exponent measure $V$, 
$$P(\eta(\bs{s}_1,t_1)\leq y_1,\eta(\bs{s}_2,t_2)\leq y_2) = \exp\left\{-V(y_1,y_2;\delta(\bs{s}_1-\bs{s}_2,t_1-t_2))\right\},$$
where in our case $V$ is given through the bivariate distribution function \eqref{limitHuesler} by
$$V(y_1,y_2;\delta(\bs{h},u)) = \frac{1}{y_1}\Phi\left(\frac{\log\frac{y_2}{y_1}}{2\sqrt{\delta(\bs{h},u)}} + \sqrt{\delta(\bs{h},u)}\right) + \frac{1}{y_2}\Phi\left(\frac{\log\frac{y_1}{y_2}}{2\sqrt{\delta(\bs{h},u)}}+ \sqrt{\delta(\bs{h},u)}\right)$$
and depends on the space and time lags $\bs{h}$ and $u$. 

In the bivariate case, the Pickands dependence function is defined through 
$$\exp\left\{-V(y_1,y_2,\delta(\bs{h},u))\right\} = \exp\left\{-\left(\frac{1}{y_1}+\frac{1}{y_2}\right)A\left(\frac{y_1}{y_1+y_2}\right)\right\}.$$
Setting $\la=y_1/(y_1+y_2)$, hence $1-\la=y_2/(y_1+y_2)$, we obtain
\begin{align*}
A(\lambda;\delta(\bs{h},u)) &= \lambda(1-\lambda)V(\lambda,1-\lambda;\delta(\bs{h},u)) \\
&= \lambda\Phi\left(\frac{\log\frac{\lambda}{1-\lambda}}{2\sqrt{\delta(\bs{h},u)}} + \sqrt{\delta(\bs{h},u)}\right) + (1-\lambda)\Phi\left(\frac{\log\frac{1-\lambda}{\lambda}}{2\sqrt{\delta(\bs{h},u)}}+ \sqrt{\delta(\bs{h},u)}\right).
\end{align*}

A useful summary measure for extremal dependence is the tail-dependence coefficient, which goes back to Geffroy~\cite{Geffroy} and Sibuya~ \cite{Sibuya}. It is defined by
$$\chi = \lim_{x \to \infty} P\left(\eta(\bs{s}_1,t_1)> F^{\leftarrow}_{\eta(\bs{s}_1,t_1)}(x) \mid \eta(\bs{s}_2,t_2)> F^{\leftarrow}_{\eta(\bs{s}_2,t_2)}(x) \right),$$
where $F^{\leftarrow}_{\eta(\bs{s},t)}$ is the generalized inverse of the marginal distribution for fixed location $\bs{s}\in S$ and time point $t\in T$. For our model this leads to 
\begin{equation}
\chi(\bs{h},u) = 2(1-\Phi(\sqrt{\delta(\bs{h},u)})).
\label{taildep}
\end{equation}
The tail dependence coefficient is a special case of the extremogram introduced in Davis and Mikosch~\cite{Davis2} (Section 1.4), with the sets $A$ and $B$ defined as $(1,\infty)$. 
The two cases $\chi(\bs{h},u) = 0$ and $\chi(\bs{h},u)=1$ correspond to the boundary cases of asymptotic independence and complete dependence. Thus, if $\delta(\bs{h},u) \to 0$, the marginal components in the bivariate case are completely dependent and if $\delta(\bs{h},u)\to \infty$, the components become independent. 
In the following section, we examine the relationship between the underlying correlation function and the tail dependence coefficient. 

\section{Possible correlation functions for the underlying space-time Gaussian process} \label{Posscov}
Provided the correlations function of the underlying Gaussian process is sufficiently smooth near $(\bs{0},0)$, then Assumption \ref{asscov} holds for some sequences $s_n$ and $t_n$. One such condition is given below. 
Throughout this section let $\bs{h} = \bs{s}_1-\bs{s}_2$ denote the space lag and $u=t_1-t_2$ the time lag.

\begin{assumption}\label{asscov2}
Assume that the correlation function allows for the following expansion 
\begin{equation*}
\rho(\bs{h},u) = 1-C_1\|\bs{h}\|^{\alpha_1} - C_2|u|^{\alpha_2} + O(\|\bs{h}\|^{\alpha_1}|u|^{\alpha_2})
\end{equation*} 
around $(\bs{0},0)$, where $0<\alpha_1,\alpha_2 \leq 2$ and $C_1,C_2\geq0$ are constants independent of $\bs{h}$ and $u$. 
\end{assumption}
\begin{remark}
The parameters $\alpha_1$ and $\alpha_2$ relate to the smoothness of the sample paths in the underlying space-time Gaussian random field.
\end{remark}
Under Assumption \ref{asscov2}, the scaling sequences in Assumption \ref{asscov} can be chosen as $s_n = (\log n)^{1/\alpha_1}$ and $t_n = (\log n)^{1/\alpha_2}$. It follows that 
\begin{equation}\label{deltaAss}
\log n (1-\rho(s_n\bs{h},t_n u)) \to C_1 \|\bs{h}\|^{\alpha_1} + C_2 |u|^{\alpha_2} = \delta(\bs{h},u), \ \text{as } n\to \infty.
\end{equation}
The condition for the tightness in \eqref{metric} can be obtained by setting 
$$D((\bs{s}_1,t_1),(\bs{s}_2,t_2)) = \max\left\{\|\bs{s}_1-\bs{s}_2\|^{\alpha_1/2},|t_1-t_2|^{\alpha_2/2}\right\},$$
which is a metric in $\bbr^d\times[0,\infty)$, since $\alpha_1,\alpha_2 \in (0,2]$. 	

Smith's storm profile model can recover a subset of the class of correlation functions specified in Assumption \ref{asscov2}. Choosing 
$\alpha_1 = \alpha_2= 2$ and
$$\sigma_{12}=0, \ \sigma_1^2 = \sigma_2^2 = \frac{1}{4C_1}, \text{  and  } \sigma_3^2 = \frac{1}{4C_2}$$
in the Smith model, we find that $\delta(\bs{h},u)$ is of the form given in Assumption \ref{asscov2}, 
$$\delta(\bs{h},u) = \frac{1}{4(\sigma_1^2\sigma_2^2 - \sigma_{12}^2)} (\sigma_2^2h_1^2 - 2\sigma_{12}h_1h_2 + \sigma_1^2h_2^2) + \frac{1}{4\sigma_3^2}u^2,$$ 
and, hence, has the same finite-dimensional distributions. 

In the following, we analyse several correlation models used in the literature for modelling Gaussian random fields in space and time.
In recent years, the interest in spatio-temporal correlation models has been growing significantly; especially in the construction of valid covariance functions in space and time. 
A simple way to construct such a model is to take the product of a spatial correlation function $\rho_1(\bs{h})$ and a temporal correlation function  $\rho_2(u)$, i.e. $\rho(\bs{h},u) = \rho_1(\bs{h})\rho_2(u)$ (see for example Cressie and Huang~\cite{Cressie}). Such a model is called separable and Assumption \ref{asscov2} is satisfied, if the spatial and the temporal correlation functions have expansions around zero of the form
$$\rho_1(\bs{h}) = 1-C_1\|\bs{h}\|^{\alpha_1} + O(\|\bs{h}\|^{\alpha_1}), \quad \rho_2(u) = 1-C_2|u|^{\alpha_2} + O(|u|^{\alpha_2}),$$  
respectively. 

\begin{example}
A more sophisticated method to obtain covariance models is given on a process-based level. 
An interesting example in this context is presented in Baxevani, Podg\'{o}rski and Rychlik {}\cite{Baxevani2, Baxevani}, who construct spatio-temporal Gaussian random fields in a continuous setup using moving averages of spatial random fields over time, given by 
$$X(\bs{s},t) = \int\limits_{-\infty}^{\infty} f(t-u) \Phi(\bs{s},du),$$
where $\Phi(\cdot,du)$ is a Gaussian random field - valued measure and $f$ is a deterministic kernel function. Using the kernel function $f(t) = e^{-\lambda t}\mathds{1}_{\left\{t\geq0\right\}}$ and the stationary spatial covariance model $\rho_1(\bs{h}) = \exp\left\{-\|\bs{h}\|^2/C\right\}$, one obtains the
separable covariance function
\begin{align*}
\gamma(\bs{h},u) &= \rho_1(\bs{h})\int\limits_{-\infty}^\infty e^{-\lambda (|u|-y)}\mathds{1}_{\left\{|u|-y\geq 0\right\}}e^{\lambda y}\mathds{1}_{\left\{y\leq0\right\}}dy = 
\rho_1(\bs{h})\int\limits_{-\infty}^{0}e^{-\lambda|u|+2\lambda y}dy \\
&= \rho_1(\bs{h})\frac{1}{2\lambda}e^{-\lambda |u|} = \frac{1}{2\lambda}\exp\left\{-\frac{\|\bs{h}\|^2}{C} -\lambda |u|\right\},
\end{align*}
where the temporal dependence is of Ornstein-Uhlenbeck type (see Example 3 in {}\cite{Baxevani2}). 
The corresponding correlation function satisfies 
$$\rho(\bs{h},u) = 1 - \frac{1}{C}\|\bs{h}\|^2 - \lambda |u| + O(\|\bs{h}\|^2|u|).$$
\end{example}
Separable space-time models do not allow for any interaction between space and time. Disadvantages of this assumption are pointed out for example in Cressie and Huang~\cite{Cressie}. Therefore, nonseparable model constructions have been developed. 
Another approach for combining purely spatial and temporal covariance functions leading also to nonseparable covariance models is introduced in Ma {}\cite{Ma2,Ma3}, given in terms of correlation functions by
$$\rho(\bs{h},u) = \int\limits_{0}^\infty\!\!\int\limits_0^{\infty} \rho_1(\bs{h}v_1)\rho_2(uv_2)dG(v_1,v_2),$$
where $G$ is a bivariate distribution function on $[0,\infty)\times [0,\infty)$.
Using the expansions above, it follows that
$$\rho(\bs{h},u) = 1- C_1\int\limits_{0}^{\infty}v_1^{\alpha_1}dG_{1}(v_1)\|\bs{h}\|^{\alpha_1} - C_2\int\limits_{0}^{\infty}v_2^{\alpha_2}dG_{2}(v_2)|u|^{\alpha_2} + O(\|\bs{h}\|^{\alpha_1}|u|^{\alpha_2}),$$
where $G_1$ and $G_2$ denote the marginal distributions of $G$, respectively. From this representation, the components in Assumption \ref{asscov2} can be defined directly.

\subsection{Gneiting's class of correlation functions}
A more elaborate class of nonseparable, stationary correlation functions is given by Gneiting's class \cite{Gneiting}.
The class of covariance functions is based on completely monotone functions, which are defined as functions $\varphi$ on $(0,\infty)$ with existing derivatives of all orders $\varphi^{(n)}, \ n =0,1,\ldots$ and 
$$(-1)^n \varphi^{(n)}(t) \geq 0,\quad t >0, \ n =0,1,\ldots.$$ 
For our purpose we use a slightly different definition of Gneiting's class. 

\begin{definition}[Gneiting's class of correlation functions \cite{Gneiting}] \label{gneiting}
Let $\varphi: \bbr_+\to \bbr$ be completely monotone and let $\psi: \bbr_+\to \bbr$ be a positive function with completely monotone derivative. Further assume that $\psi(0)^{-d/2}\varphi(0) = 1$, where $d$ is the spatial dimension, and $\beta_1,\beta_2\in (0,1]$. 
The function 
$$\rho(\bs{h},u) = \frac{1}{\psi\left(\left|u\right|^{2\beta_2}\right)^{d/2}}\varphi\left(\frac{\left\|\bs{h}\right\|^{2\beta_1}}{\psi\left(\left|u\right|^{2\beta_2}\right)}\right), \quad (\bs{h},u) \in \bbr^d \times \bbr_+ ,$$
defines a non-separable, isotropic space-time correlation function with $\rho(\bs{0},0) = 1$.
\end{definition}

Compared to the original definition in \cite{Gneiting}, we included the parameters $\beta_1$ and $\beta_2$, which is not a restriction since we can simply change the norms by defining $\|\cdot\|_{*}$ and $|\cdot|_{*}$ in terms of the old ones through
$$\|\bs{h}\|_{*} = \|\bs{h}\|^{\beta_1} , \quad \text{and} \quad |u|_* = |u|^{\beta_2}.$$
These new quantities are still norms since $\beta_1,\beta_2 \in (0,1]$.
In the next step, we provide an expansion of the correlation function around zero to obtain Assumption \ref{asscov2}. The following proposition generalizes a result by Xue and Xiao~\cite{Xue} (Proposition 6.1).   
\begin{proposition}
Assume that $\psi'(0) \neq 0$.
The correlation function taken from the Gneiting class satisfies Assumption \ref{asscov2} with $\alpha_1 = 2\beta_1$, $\alpha_2 = 2\beta_2$ and 
\begin{equation}
C_1 = \psi(0)^{-d/2}\left(\int\limits_{0}^{\infty}zdF_{\varphi}(z)\bigg{/}\int\limits_{0}^{\infty}dF_{\varphi}(z)\right), \quad 
C_2 = \frac{d}{2}\psi(0)^{-1}\psi'(0),
\label{C1C2Gneiting}
\end{equation}
where $F_{\varphi}$ is a non-descreasing bounded function with $F_{\varphi}(0)\neq 0$ and $\int_0^{\infty} z dF_{\varphi}(z) <\infty$.  
\end{proposition}
\begin{proof}
Since the function $\varphi$ is completely monotone, Bernstein's theorem (see for example Feller~\cite{Feller}, Chapter 13) gives 
$$\varphi(x) = \int\limits_{0}^{\infty}e^{-xz}dF_{\varphi}(z), \quad x\geq0.$$
From the properties of the correlation function $\psi(0)^{-d/2}\varphi(0) = 1$, it follows that $\psi(0)\neq 0$ and $\varphi(0) \neq 0$. 
We apply a Taylor expansion to the functions $\psi(\cdot)^{-d/2}$ and the exponential in the representation of $\varphi$: 
\begin{align*}
\psi(u)^{-d/2} &= \psi(0)^{-d/2} -\frac{d}{2}\psi(0)^{-d/2-1}\psi'(0)u + o(u),\ \ u\to0 \\
\varphi(x) &= \int\limits_{0}^{\infty}(1-xz + o(x)) dF_{\varphi}(z) = \int\limits_{0}^{\infty}dF_{\varphi}(z) - x\int\limits_{0}^{\infty}zdF_{\varphi}(z) + o(x), \ \ x\to 0.
\end{align*}
Using the expansions in the correlation function and replacing $u$ by $|u|^{2\beta_2}$ and $x$ by $\|\bs{h}\|^{2\beta_1}/\psi(|u|^{2\beta_2})$, we obtain 
\begin{align*}
\rho(\bs{h},u) &= \left(\psi(0)^{-d/2} - \frac{d}{2}\psi(0)^{-d/2-1}\psi'(0)|u|^{2\beta_2} + o(|u|^{2\beta_2})\right) \\
&\times \left(\int\limits_{0}^{\infty}dF_{\varphi}(z) - \int\limits_{0}^{\infty}zdF_{\varphi}(z)\|\bs{h}\|^{2\beta_1} \left[\psi(0)^{-d/2} - \frac{d}{2} \psi(0)^{-d/2-1}\psi'(0)|u|^{2\beta_2} +o(|u|^{2\beta_2})\right] + o(\|\bs{h}\|^{2\beta_1}) \right)\\
&= \psi(0)^{-d/2}\varphi(0) - \frac{d}{2}\psi(0)^{-d/2-1}\varphi(0)\psi'(0)|u|^{2\beta_2} - \psi(0)^{-d}\int\limits_{0}^{\infty}zdF_{\varphi}(z)\|\bs{h}\|^{2\beta_1} + O(\|\bs{h}\|^{2\beta_1}|u|^{2\beta_2}) \\
&= 1 - \frac{d}{2}\psi(0)^{-1}\psi'(0)|u|^{2\beta_2} - \psi(0)^{-d/2}\left(\int\limits_{0}^{\infty}zdF_{\varphi}(z)\bigg{/}\int\limits_{0}^{\infty}dF_{\varphi}(z)\right)\|\bs{h}\|^{2\beta_1} + O(\|\bs{h}\|^{2\beta_1}|u|^{2\beta_2})\\
&= 1 - C_1\|\bs{h}\|^{2\beta_1} - C_2 |u|^{2\beta_2} + O(\|\bs{h}\|^{2\beta_1}|u|^{2\beta_2}),
\end{align*}
where $C_1$ and $C_2$ are defined as in \eqref{C1C2Gneiting}. 
\end{proof}

\begin{figure}[htbp]
\centering
\includegraphics[scale=0.4]{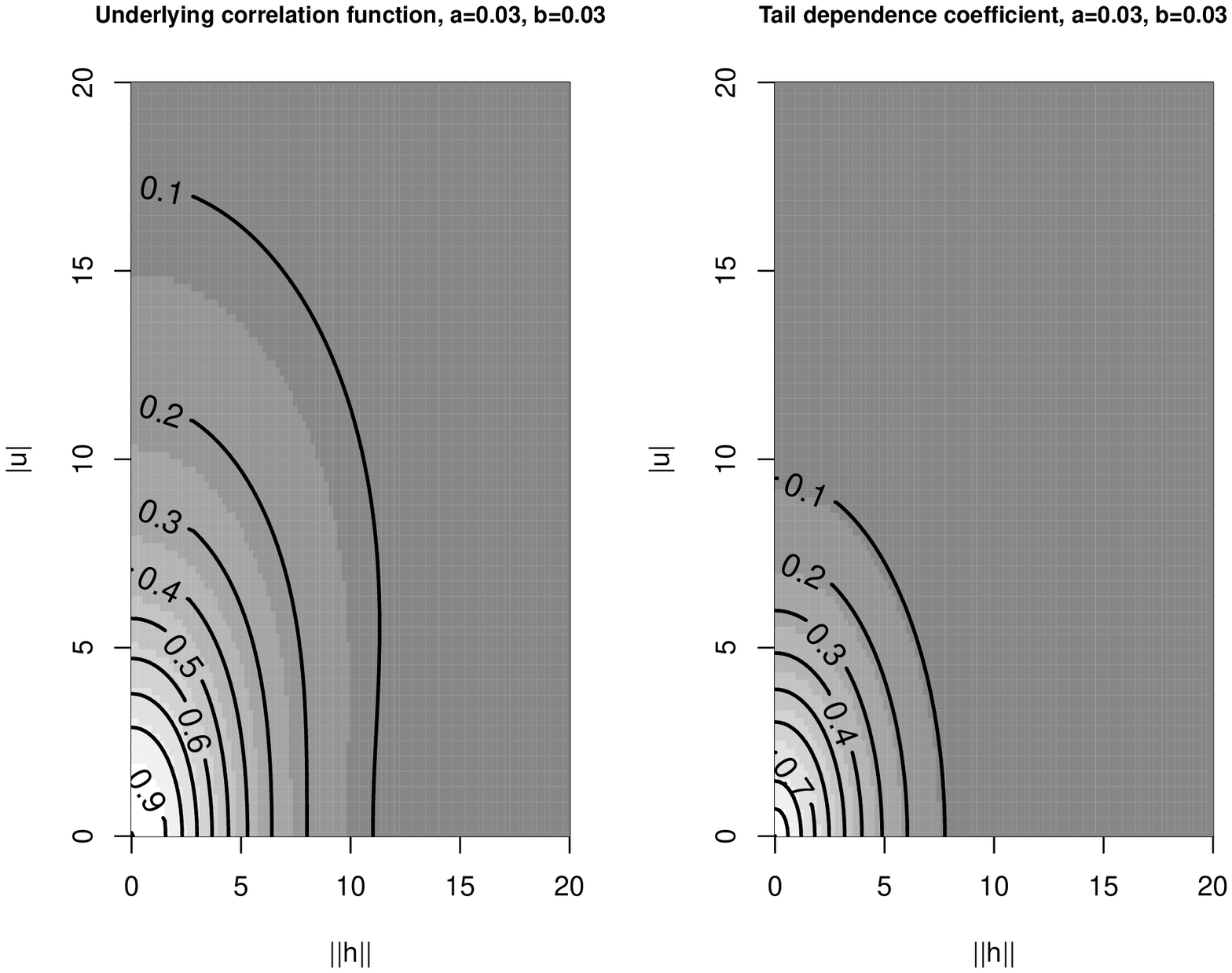}
\includegraphics[scale=0.4]{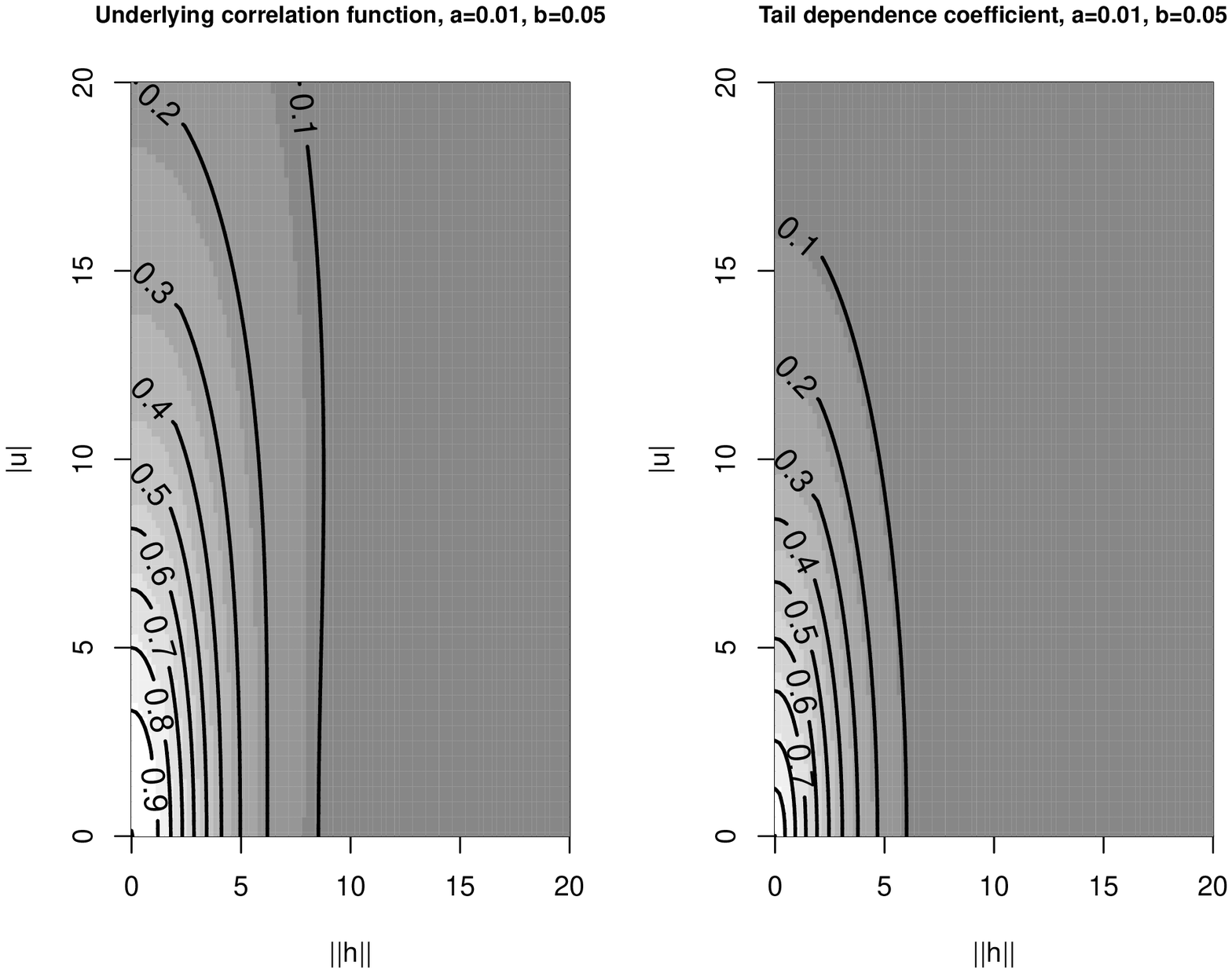}
\includegraphics[scale=0.4]{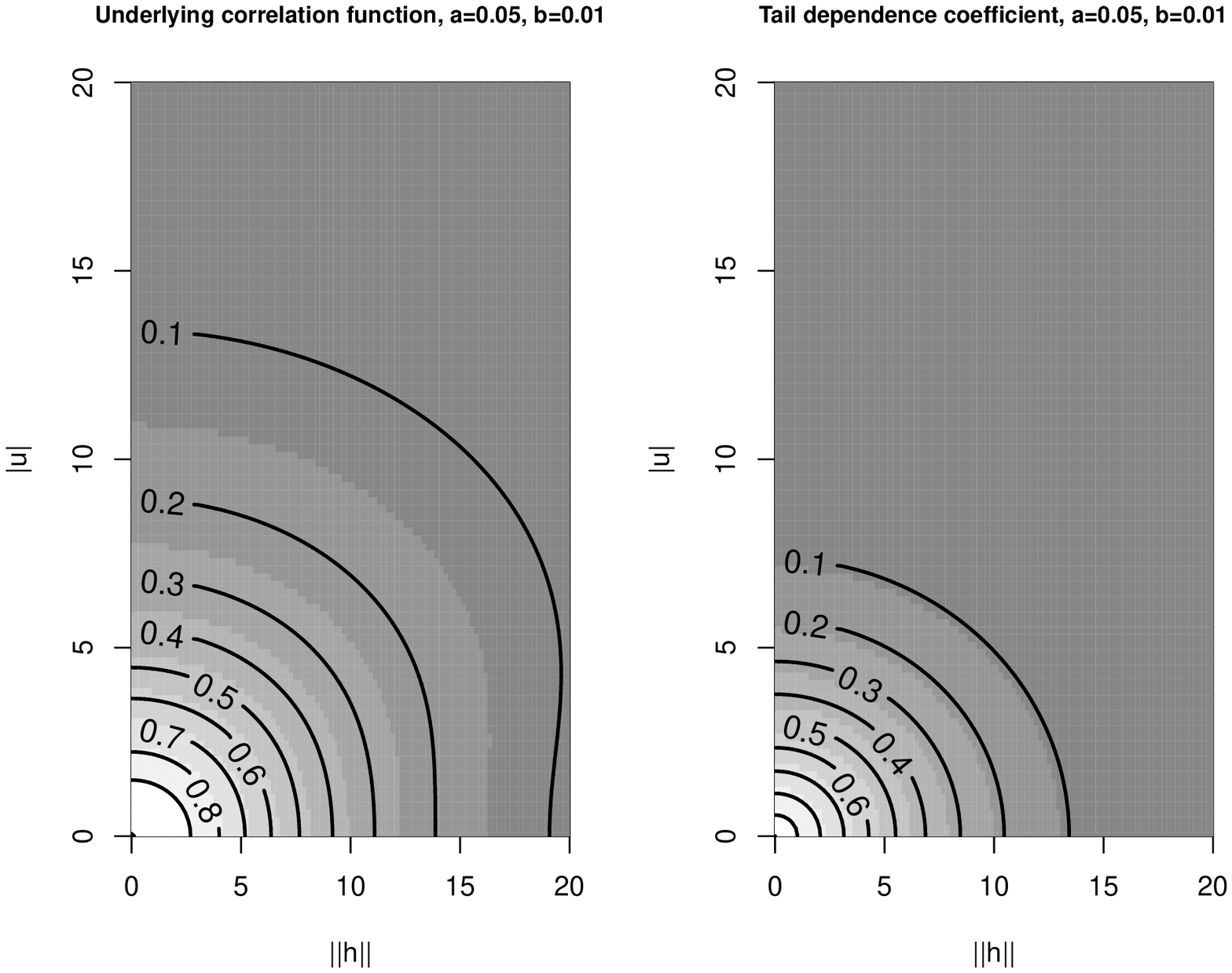}
\caption{Contour plots for the underlying correlation function (left) and the resulting tail dependence coefficient (right) depending on the absolute space lag $\|\bs{h}\|$ and time lag $|u|$ for different values of the scaling parameters $a$ (time) and $b$ (space), $\nu=3/2$ and $\gamma=1$.}
\label{covtailplot}
\end{figure}

\begin{remark}
For various choices of $\alpha_1,\alpha_2 \in (0,2]$, the correlation functions in the Gneiting class have the flexibility to model different levels of smoothness of the underlying Gaussian random fields.
\end{remark}
\begin{example}\label{ExGneiting}
We illustrate with a specific example, where the functions $\phi$ and $\psi$ are taken from \cite{Gneiting}; namely
\begin{align*}
\varphi(x) &= (1+bx)^{-\nu},   \quad \ \psi(x) = (1+ax)^{\gamma},
\end{align*} 
where $a,b,\nu > 0$ and $0<\gamma\leq 1$.
The function $\varphi$ is the Laplace transform of a gamma probability density function with shape $\nu>0$ and scale $b>0$, which has mean $b\nu$. The first-order derivative of $\psi$ at zero is given by $\psi'(0) = a\gamma$. We choose $\beta_1=\beta_2=1$ leading to $\alpha_1 = \alpha_2 = 2$ and, thus, a mean-square differentiable Gaussian random field. 
The constants $C_1$ and $C_2$ are given by 
$$C_1=b\nu, \quad \text{and  }  C_2 = \frac{d}{2}a\gamma.$$
Figure \ref{covtailplot} shows contour plots of the correlation function and the resulting tail dependence coefficient as in \eqref{taildep} based on different values for $a$ and $b$ with $\nu=3/2$ and $\gamma=1$ fixed as a function of the space-lag $\|\bs{h}\|$ and time lag $|u|$. We see that the tail dependence function exhibits virtually the identical pattern of the underlying correlation function under a compression of the space-time scale. In particular, the extremal dependence dies out more quickly for large space and time lags than for the correlation function.  

In a second step, we simulate random fields in space and time using the above defined correlation model with $a=b=0.03$, $\nu=3/2$ and $\gamma=1$. We start the simulation procedure with $n=100$ replications of a Gaussian random field with correlation function $\rho(s_n\bs{s},t_n u)$ using the simulation routine in the R-package \verb RandomFields  by Schlather~\cite{Schlather1}. The random fields are then transformed to standard Fr\'echet and the pointwise maximum is taken over the $100$ replications. Figure \ref{SimFrechet} shows image and perspective plots (using the R-package \verb fields  for visualization) of the simulated random fields for four consecutive time points. Figures \ref{SimGumbel} and \ref{SimWeibull} show the resulting random fields, if the margins are transformed to standard Gumbel and Weibull instead of Fr\'echet distributions. Clearly, in both cases the peaks are not as high as in the Fr\'echet case, leading to a smoother appearance of the resulting random field. One still sees the isolation of the peaks in the Fr\'echet case, which are well-known from the storm model of Smith using a centered Gaussian density for the function $f$. However, in the other two cases, they are not as pronounced. 
\end{example}  
  
\begin{center}
\vspace*{-1cm}
\includegraphics[scale=0.36]{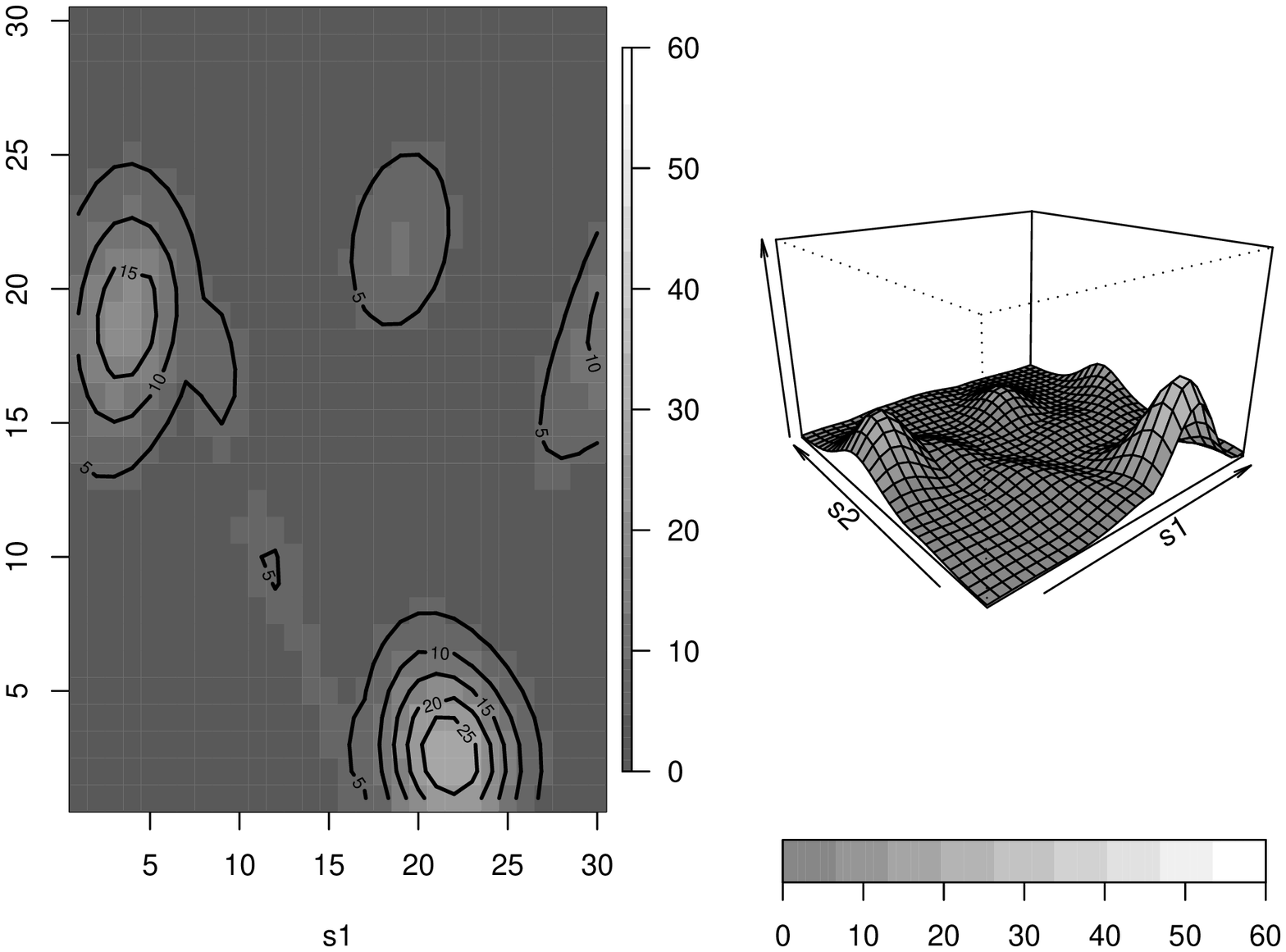}
\vspace*{-0.8cm}

\includegraphics[scale=0.36]{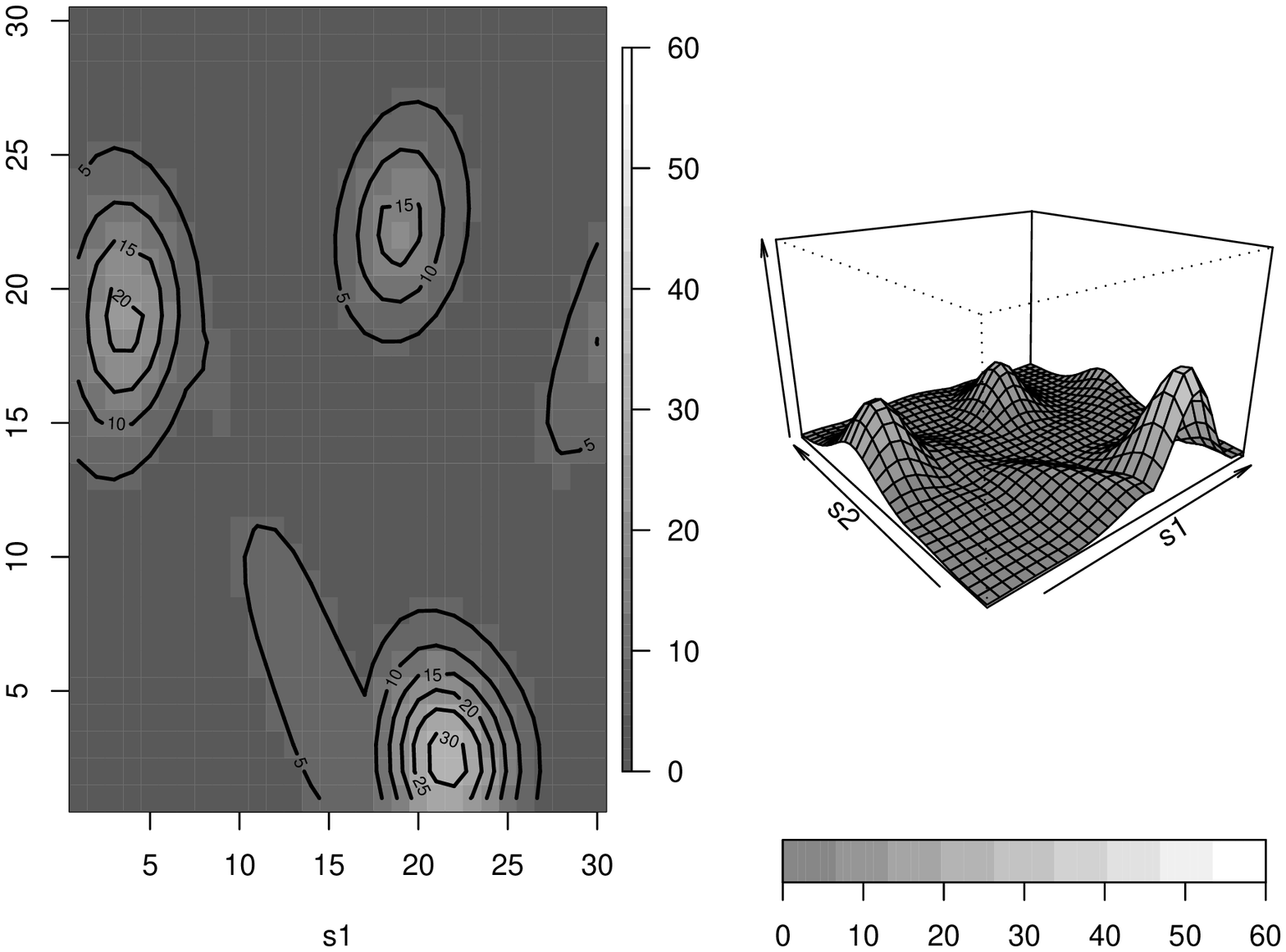}
\vspace*{-0.8cm}

\includegraphics[scale=0.36]{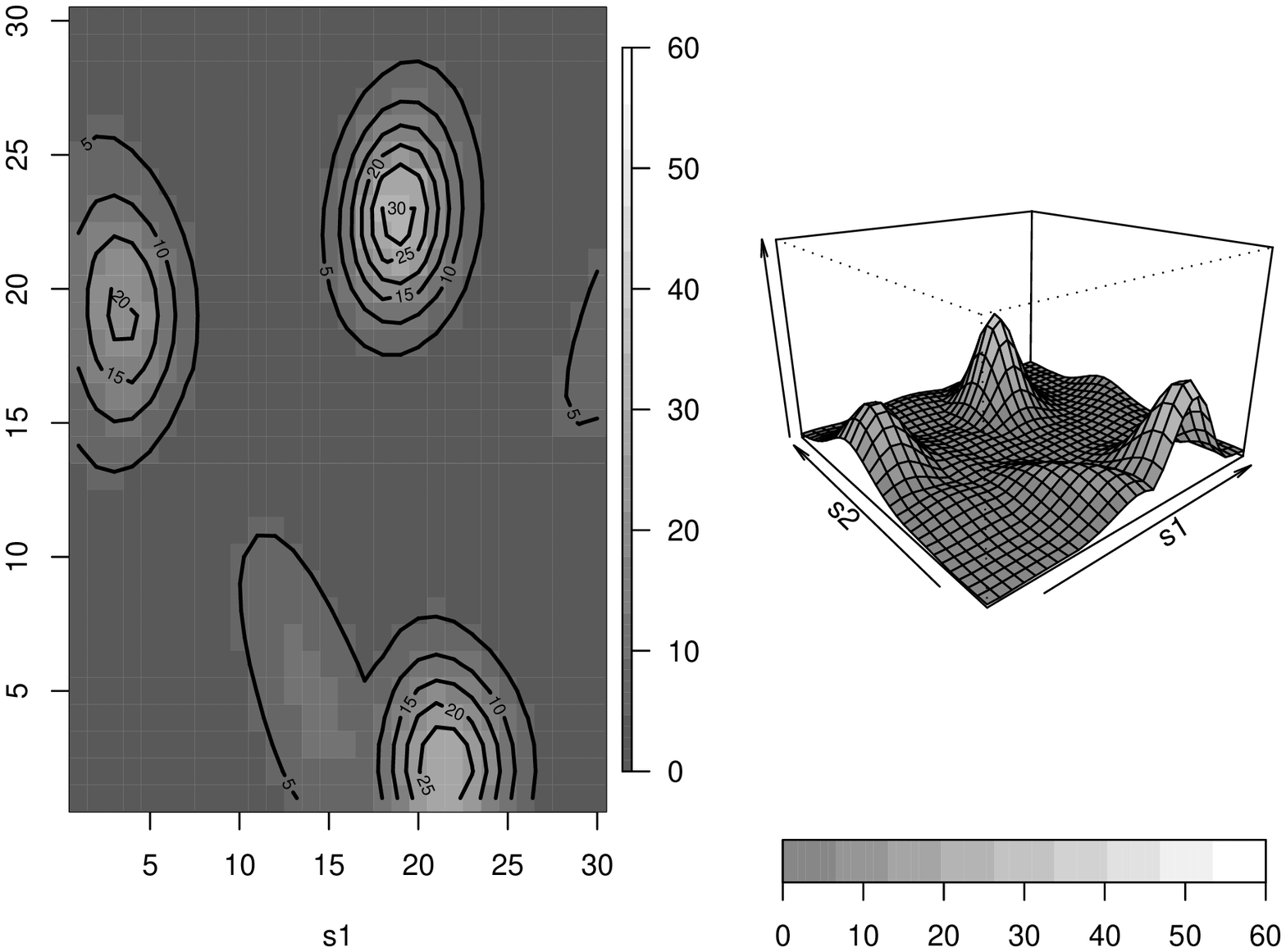}
\vspace*{-0.8cm}

\includegraphics[scale=0.36]{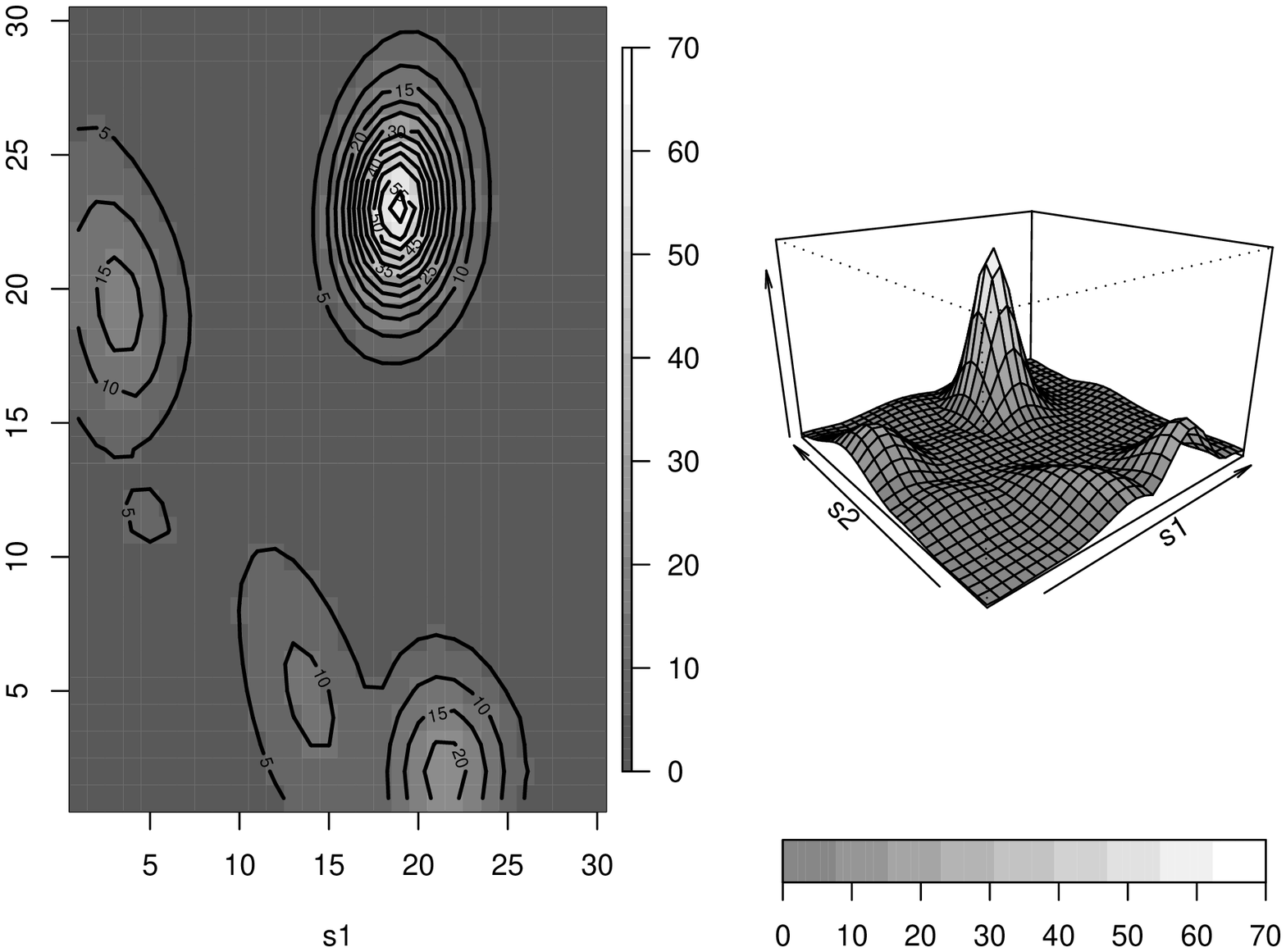}

\captionof{figure}{Simulated max-stable random fields with Fr\'{e}chet margins ($a=0.03, b=0.03, \nu=-3/2$, $\gamma=1$) for four consecutive time points (from the top to the bottom) with a time lag of one using a grid simulation of size $30 \times 30$.}
\label{SimFrechet}
\end{center}

\begin{center}
\vspace*{-1cm}
\includegraphics[scale=0.36]{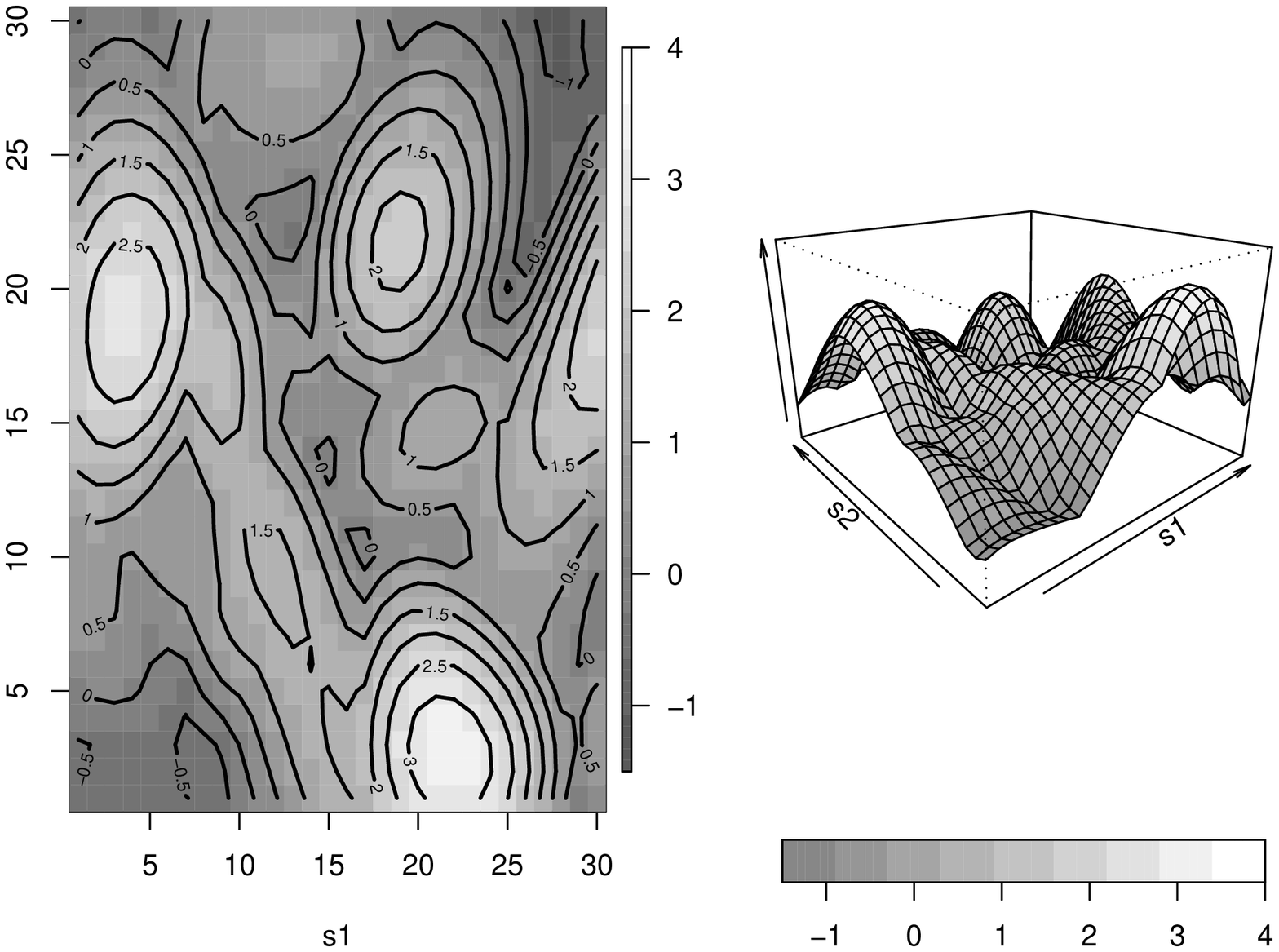}
\vspace*{-0.8cm}

\includegraphics[scale=0.36]{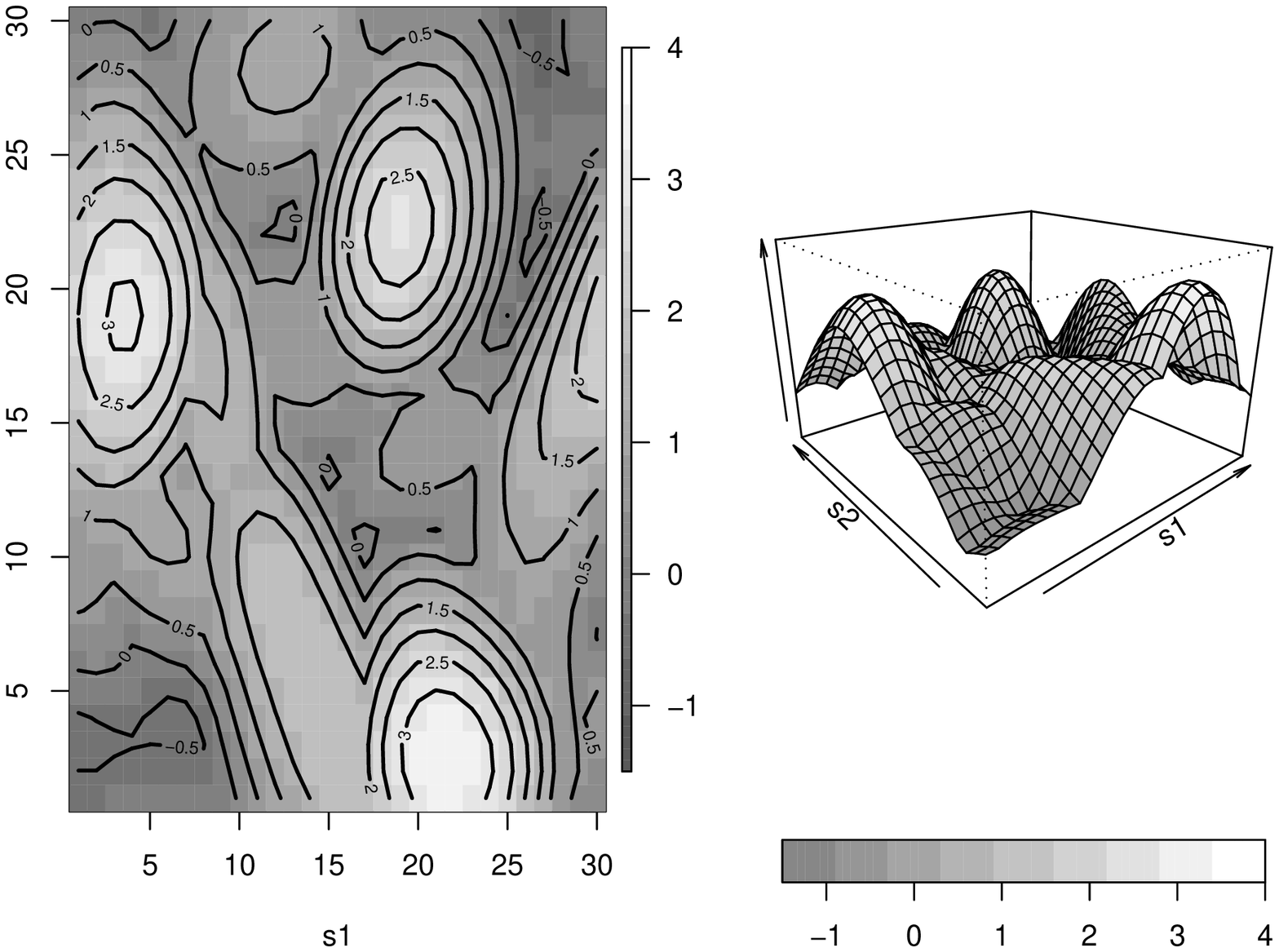}
\vspace*{-0.8cm}

\includegraphics[scale=0.36]{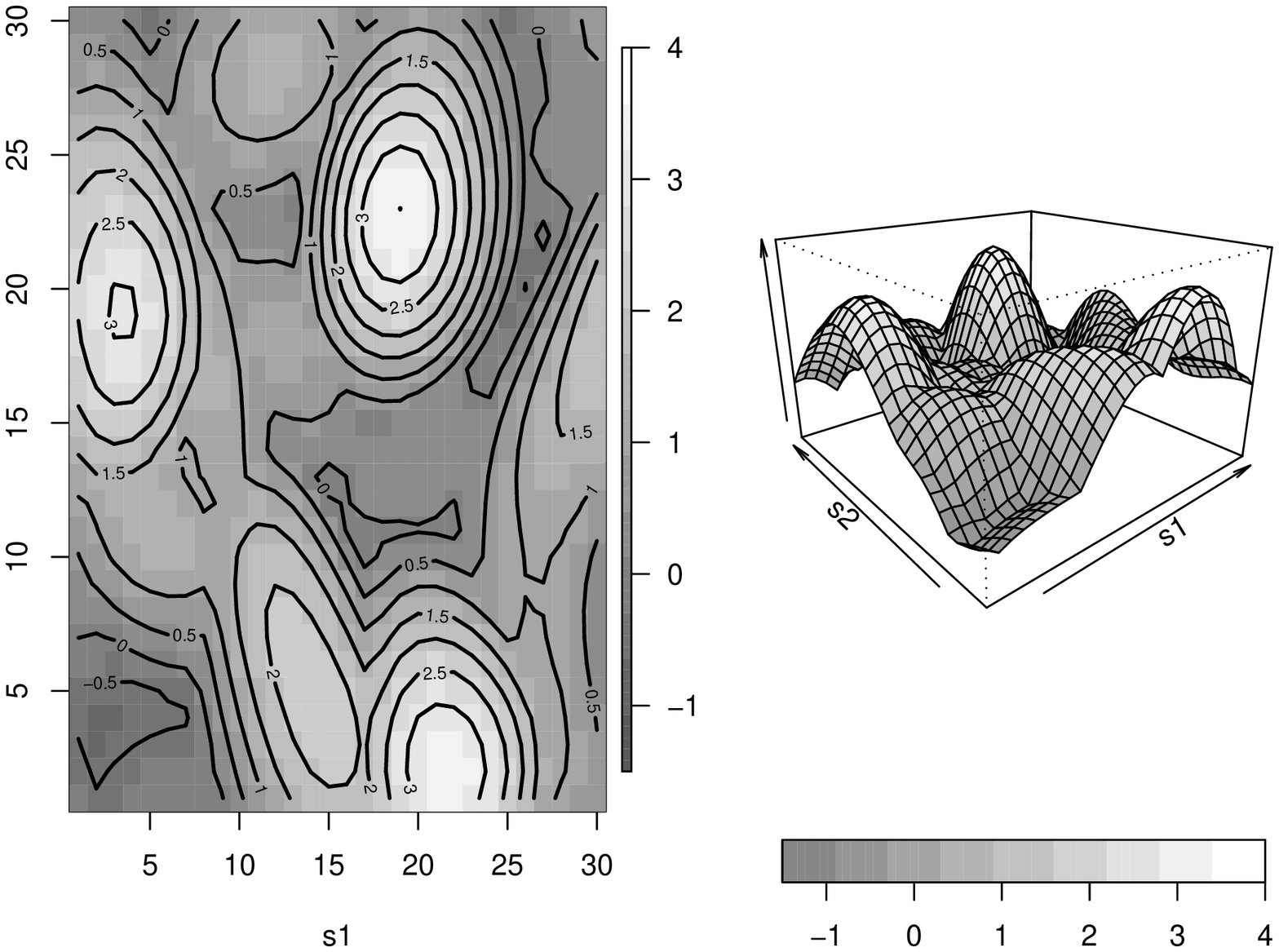}
\vspace*{-0.8cm}

\includegraphics[scale=0.36]{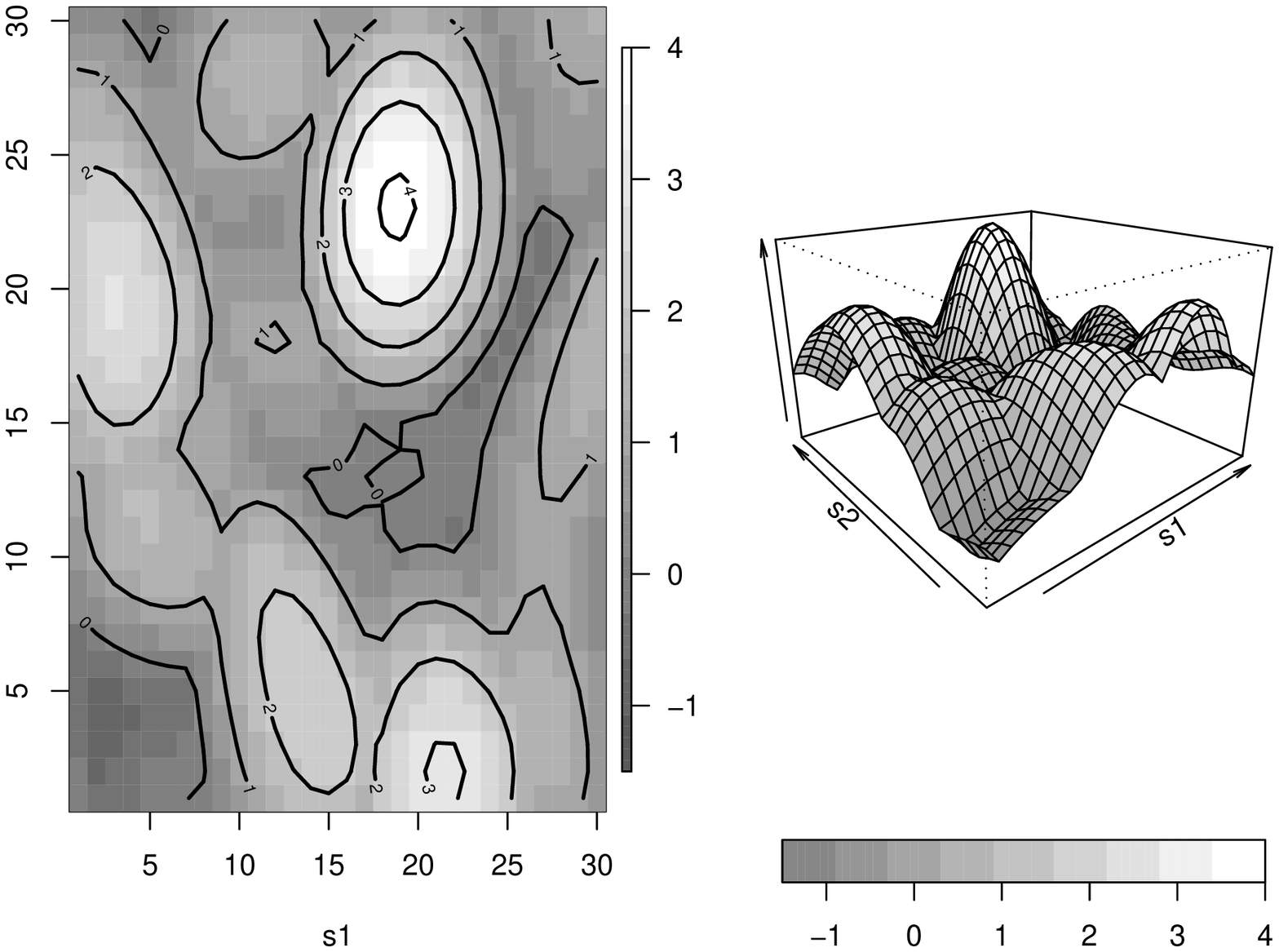}

\captionof{figure}{Simulated max-stable random fields with Gumbel margins ($a=0.03, b=0.03, \nu=-3/2$) for four consecutive time points (from the top to the bottom) with a time lag of one using a grid simulation of size $30 \times 30$.}
\label{SimGumbel}
\end{center}

\begin{center}
\vspace*{-1cm}
\includegraphics[scale=0.36]{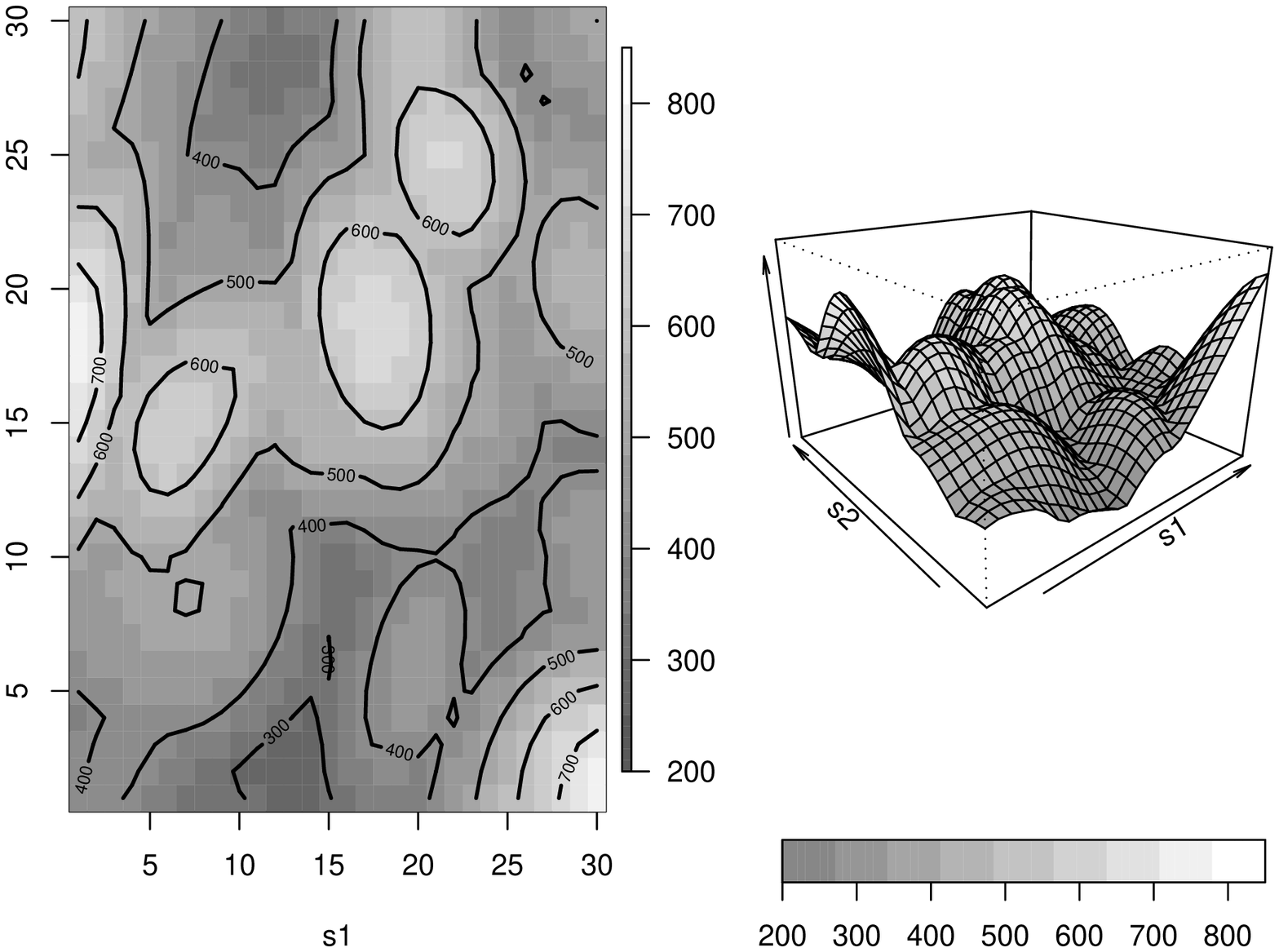}
\vspace*{-0.8cm}

\includegraphics[scale=0.36]{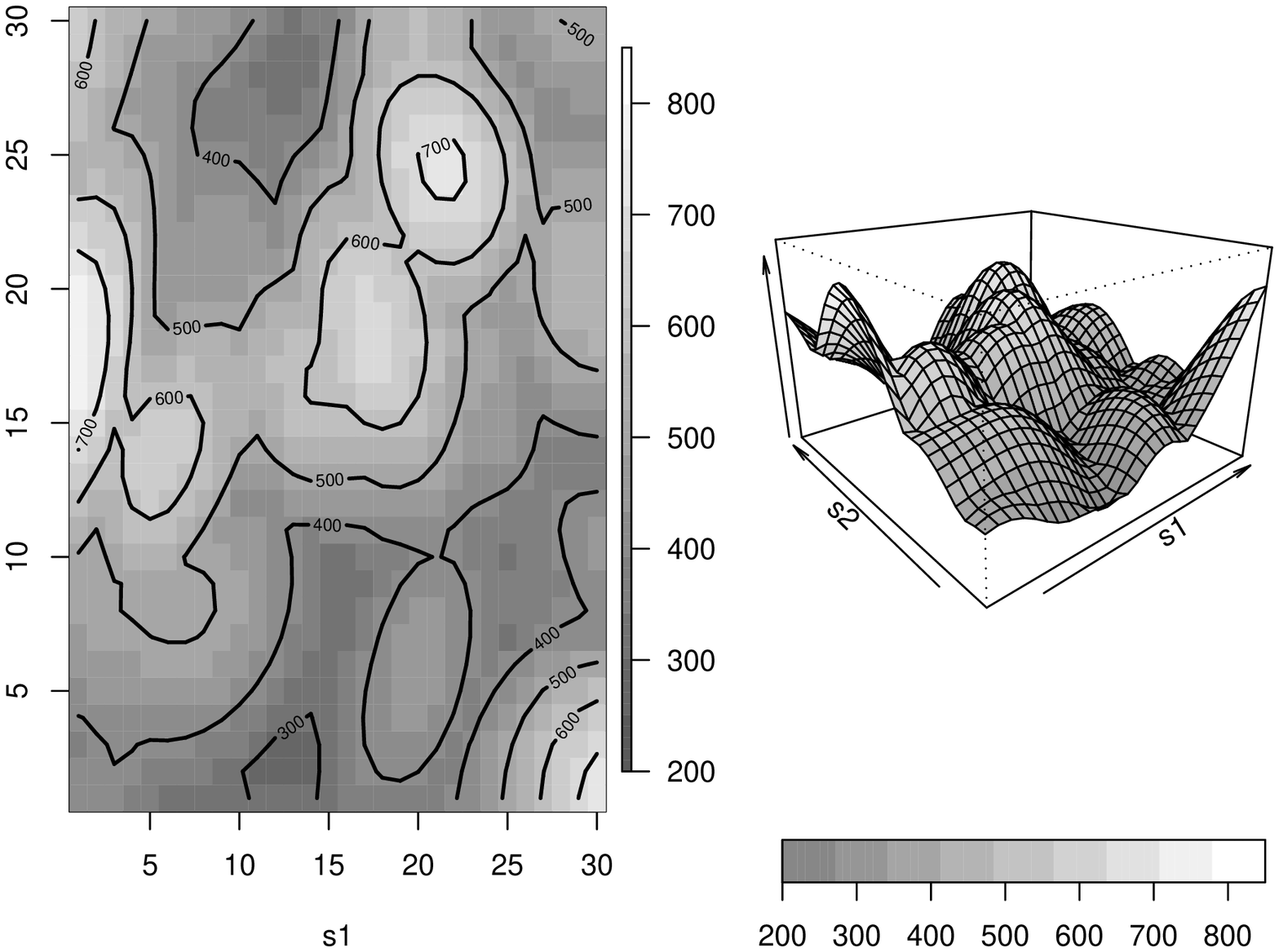}
\vspace*{-0.8cm}

\includegraphics[scale=0.36]{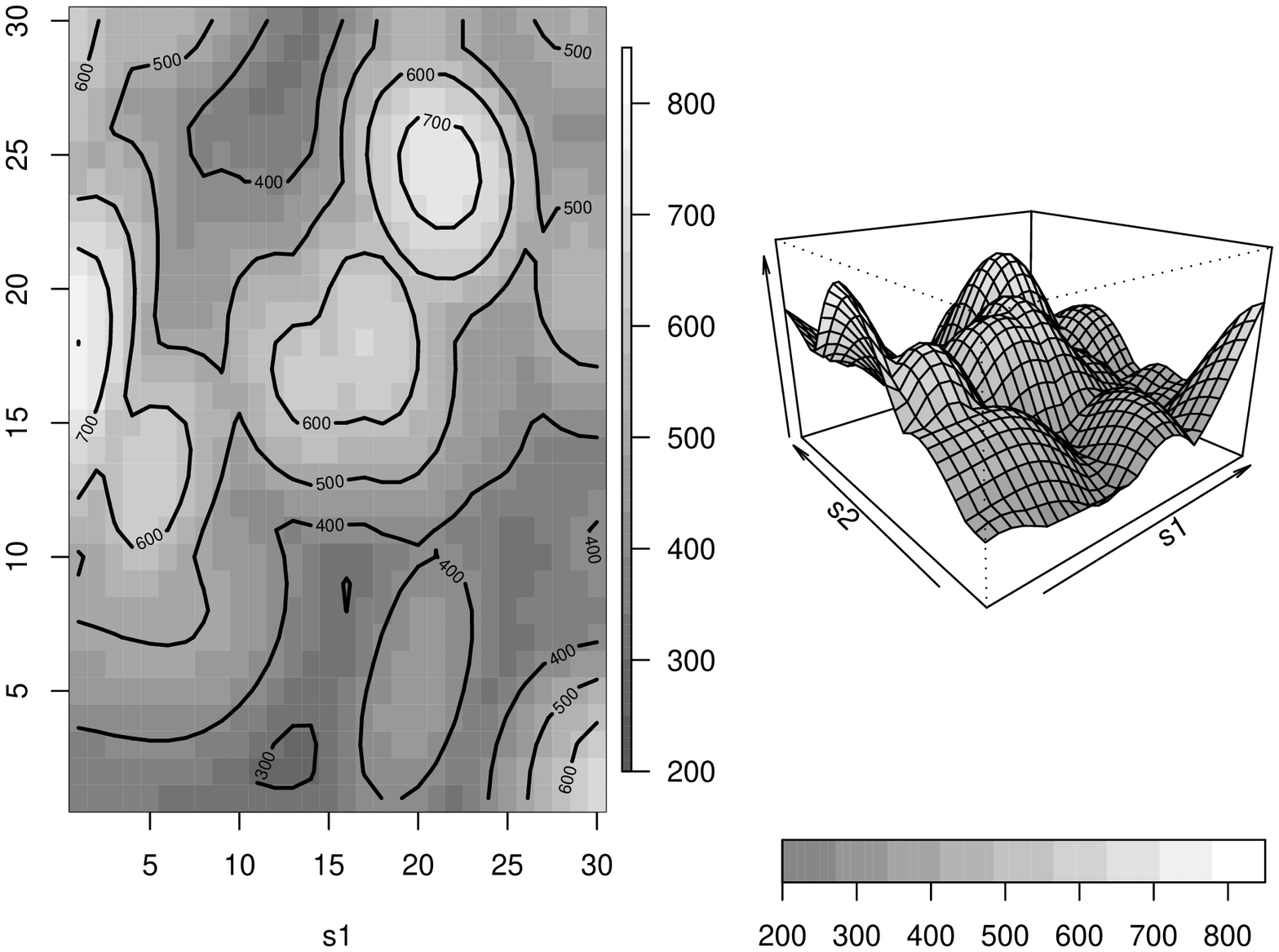}
\vspace*{-0.8cm}

\includegraphics[scale=0.36]{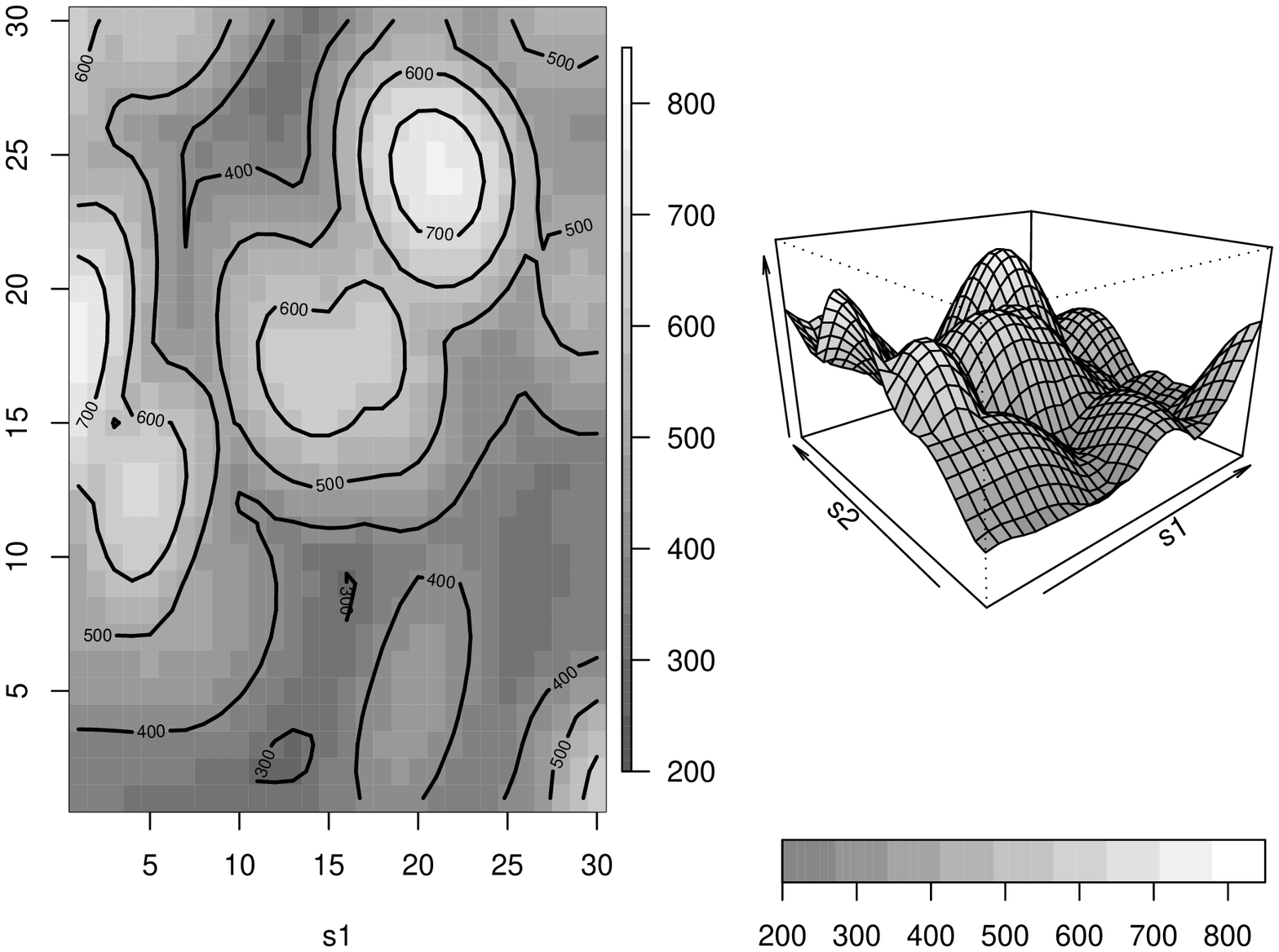}

\captionof{figure}{Simulated max-stable random fields with Weibull margins ($a=0.03, b=0.03, \nu=-3/2$) for four consecutive time points (from the top to the bottom) with a time lag of one using a grid simulation of size $30 \times 30$.}
\label{SimWeibull}
\end{center}

\subsection{Modelling spatial anisotropy}\label{aniso}
The correlation functions of the underlying Gaussian random fields in the previous sections were assumed to be spatially isotropic, meaning that the correlation function only depends on the absolute space and time lags $\|\bs{h}\|$ and $|u|$. An easy way to introduce spatial anisotropy to a model is given by geometric anisotropy, i.e., 
$$\tilde{\rho}(h,u) = \rho(\left\|Ah\right\|,\left|u\right|),$$
where $A$ is a transformation matrix.\\ 
In the two dimensional case geometric anisotropy in space can be modelled by a transformation matrix $A=TR,$ with rotation and distance matrix
where $$R = \begin{pmatrix} \cos \alpha & -\sin \alpha \\ \sin \alpha & \cos \alpha\end{pmatrix}, \quad T = \begin{pmatrix} 1/a_{\text{max}} & 0 \\ 0 & 1/a_{\text{min}}\end{pmatrix}.$$
Geometric anisotropy directly relates to the tail dependence coefficient 
$$\chi(\bs{h},u) = 2(1-\Phi(\sqrt{\delta(A\bs{h},u)})).$$
Figure \ref{geometricaniso} compares isotropic and anisotropic correlation functions and the corresponding tail dependence coefficients as function of the space lag components $\bs{h} = (h_1,h_2)'$, where the isotropic correlation is the same as in Example \ref{ExGneiting} with $\gamma=1$, $\nu=3/2$ and $a=b=0.03$. For the anisotropic case we choose $a_{\text{min}}=1$, $a_{\text{max}} = 3$ and $\alpha= 45^{\circ}$. It can be seen, that the structure in the correlation function translates to the tail dependence coefficient. Corresponding max-stable random fields with Fr\'echet margins are shown for four consecutive time points in Figure \ref{anisofields}. From the image plots, one clearly sees that the correlation is stronger in one direction. The perspective plots show that the isolated peaks are now stretched in one direction. In reality, this could correspond to wind speed peaks coming for example from a storm shaped particular in this wind direction.  
 
A more complex way of introducing anisotropy in space is given by the Bernstein class, which is introduced in Porcu et al.~\cite{Porcu} and revisited in Mateu et al.~\cite{Mateu}. 
The covariance model is defined by 
$$C(\bs{h},u) = \int\limits_{0}^{\infty}\!\!\int\limits_{0}^{\infty} \exp\left\{-\sum\limits_{i=1}^d\psi_i(|h_i|)v_1 - \psi_{t}(|u|)v_2\right\}dF(v_1,v_2), $$
where $F$ is a bivariate distribution function and $\psi_i, i=1,\ldots,d$ and $\psi_t$ are positive functions on $[0,\infty)$ with completely monotone derivatives, also called Bernstein functions. We assume that $\psi_i, i=1,\ldots,d$ and $\psi_t$ are standardized, such that $\psi_i(0)=\psi_t(0)=1, i=1,\ldots,d$. Assumption \ref{asscov} can directly be derived for the corresponding correlation function. 

\begin{align*}
\rho(\bs{h},u) ={} & C(\bs{h},u)/C(\bs{0},0) \\
={} & \bigg{(}1 - \sum\limits_{i=1}^d\psi_i(|h_i|)\int\limits_{0}^{\infty}\!\!\int\limits_{0}^{\infty} v_1dF(v_1,v_2) - \psi_{t}(|u|)\int\limits_{0}^{\infty}\!\!\int\limits_{0}^{\infty} v_2dF(v_1,v_2)\bigg{)}\bigg{/} \\
& \bigg{(}1-d\int\limits_{0}^{\infty}\!\!\int\limits_{0}^{\infty} v_1dF(v_1,v_2) - \int\limits_{0}^{\infty}\!\!\int\limits_{0}^{\infty} v_2dF(v_1,v_2)\bigg{)} \\
={} & \bigg{(}1-\sum\limits_{i=1}^d (1-C_1|h_i|^{\alpha_1} + o(|h_i|^{\alpha_1}))\int\limits_{0}^\infty v_1 F_{v_1}(v_1) - (1-C_2|u|^{\alpha_2} + O(|u|^{\alpha_2}))\int\limits_{0}^{\infty}v_2 dF_{v_2}(v_2)\bigg{)}\bigg{/} \\
& \bigg{(}1-d\int\limits_{0}^{\infty}v_1dF_{v_1}(v_1) - \int\limits_{0}^{\infty}v_2dF_{v_2}(v_2)\bigg{)} \\
={} & 1 - C_1\int\limits_{0}^{\infty}v_1dF_{v_1}(v_1)\sum\limits_{i=1}^d |h_i|^{\alpha_1} -C_2\int\limits_{0}^{\infty}v_2 dF_{v_2}(v_2)|u|^{\alpha_2} + O(\sum\limits_{i=1}^d |h_i|^{\alpha_1}) + O(|u|^{\alpha_2})
\end{align*}

\begin{center}
\includegraphics[scale=0.36]{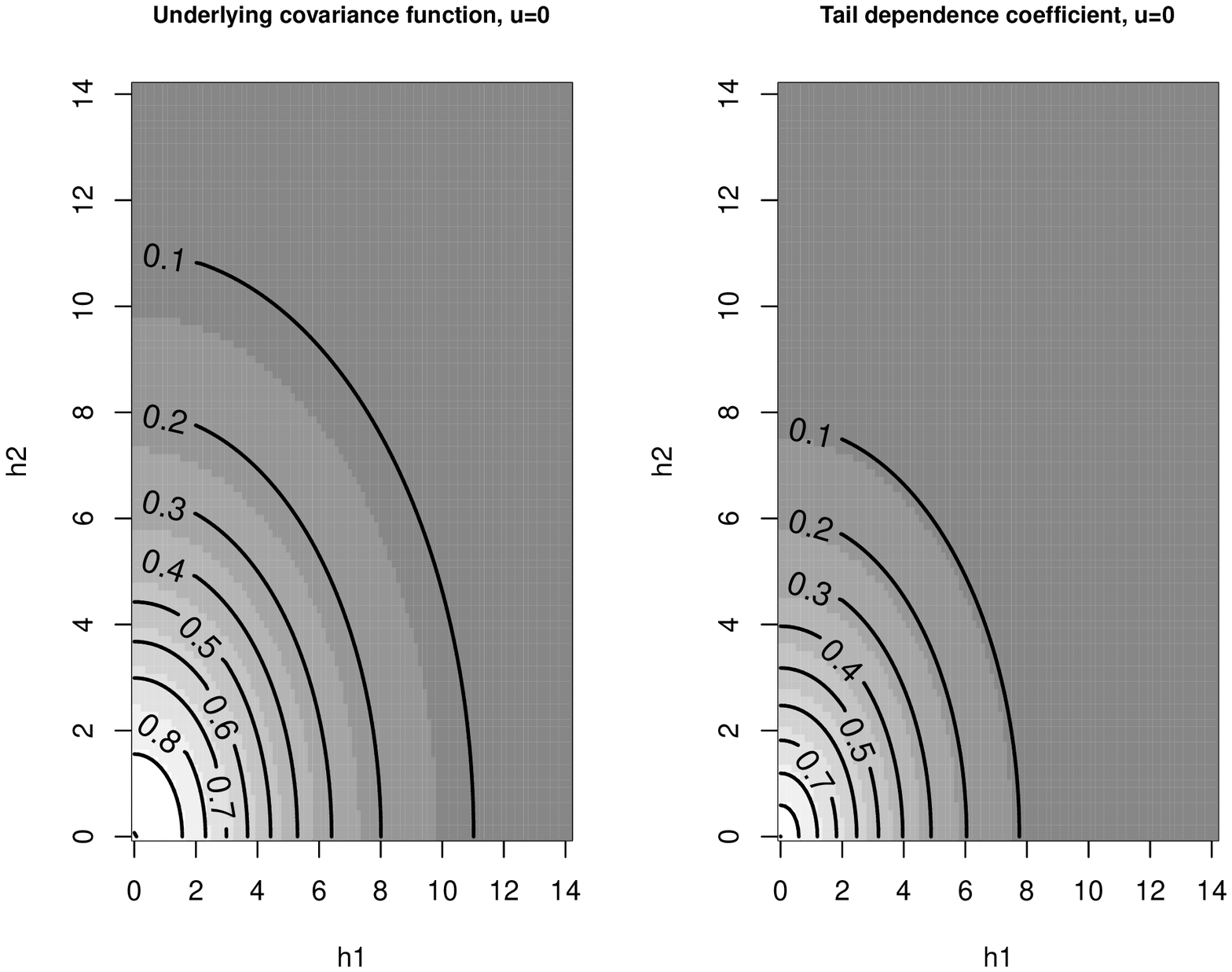}
\includegraphics[scale=0.36]{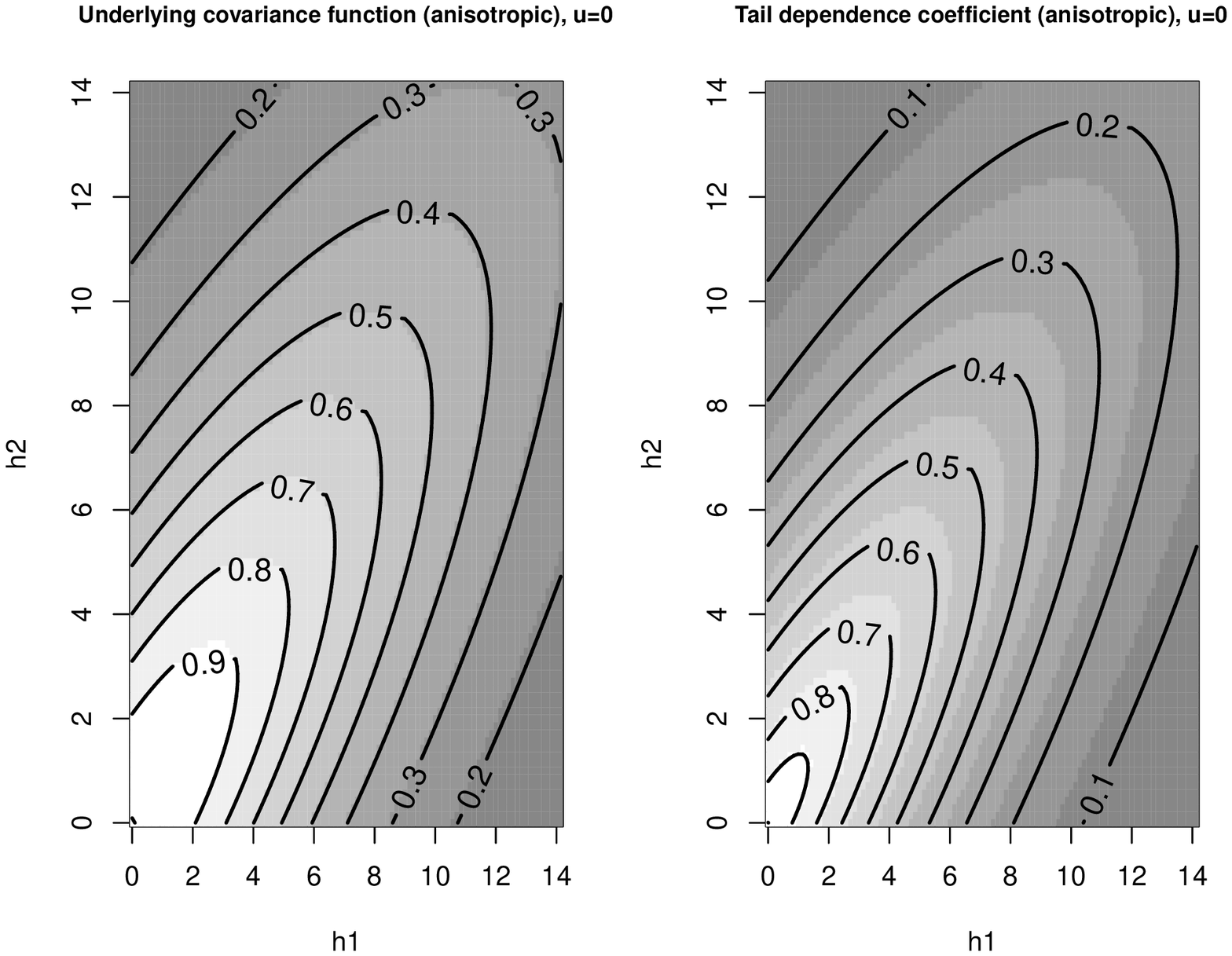}
\captionof{figure}{Contour plots for covariance functions and tail dependence coefficients depending on the space lag components $h_1$ and $h_2$ in the isotropic case (top) and for included geometric anisotropy (bottom) for $a=b=0.03$, $a_{\text{min}}=1$, $a_{\text{max}}=2$ and $\alpha=45^{\circ}$. }
\label{geometricaniso}
\end{center}

\newpage
\begin{center}
\vspace*{-1cm}
\includegraphics[scale=0.36]{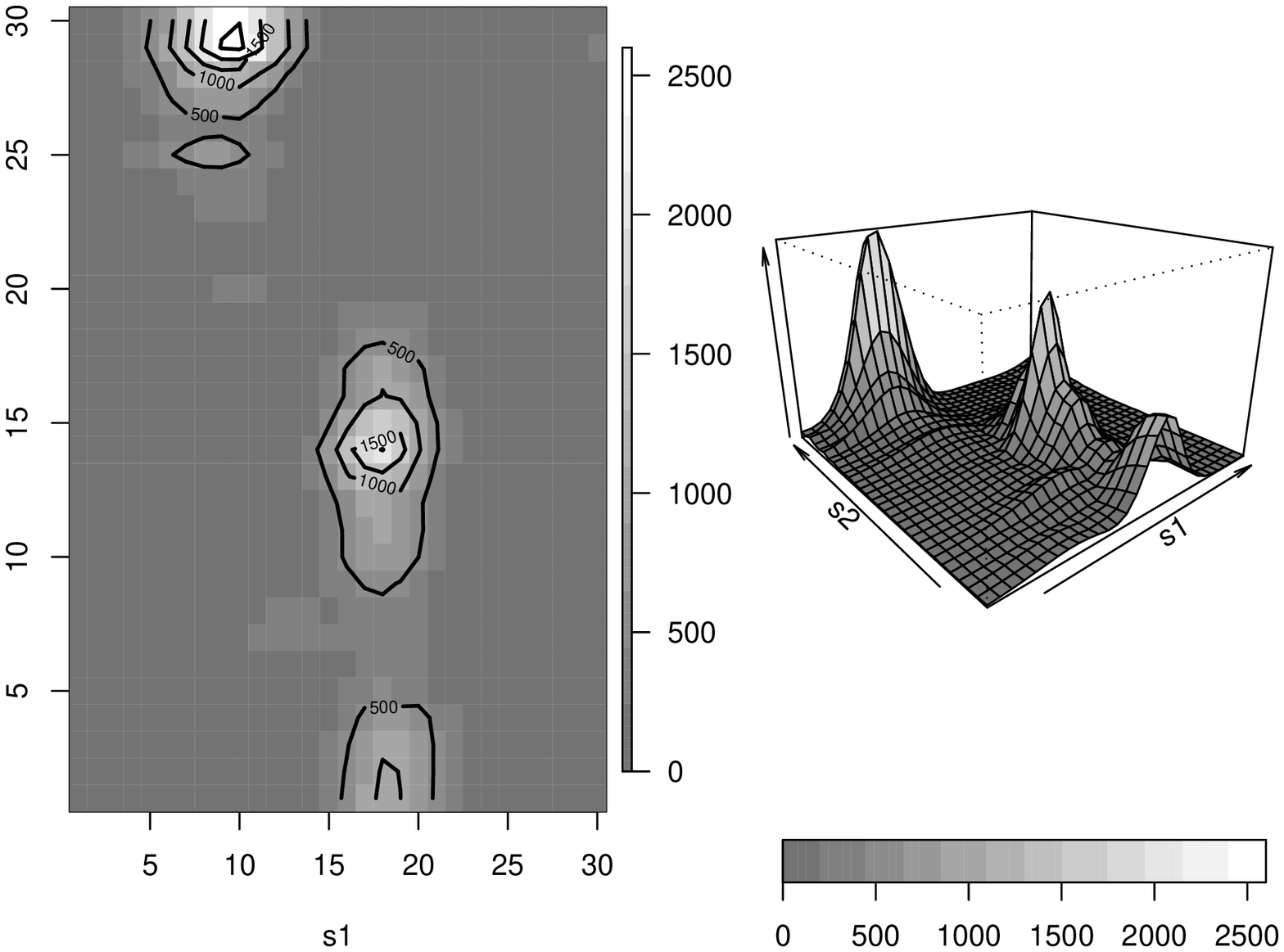}
\vspace*{-0.8cm}

\includegraphics[scale=0.36]{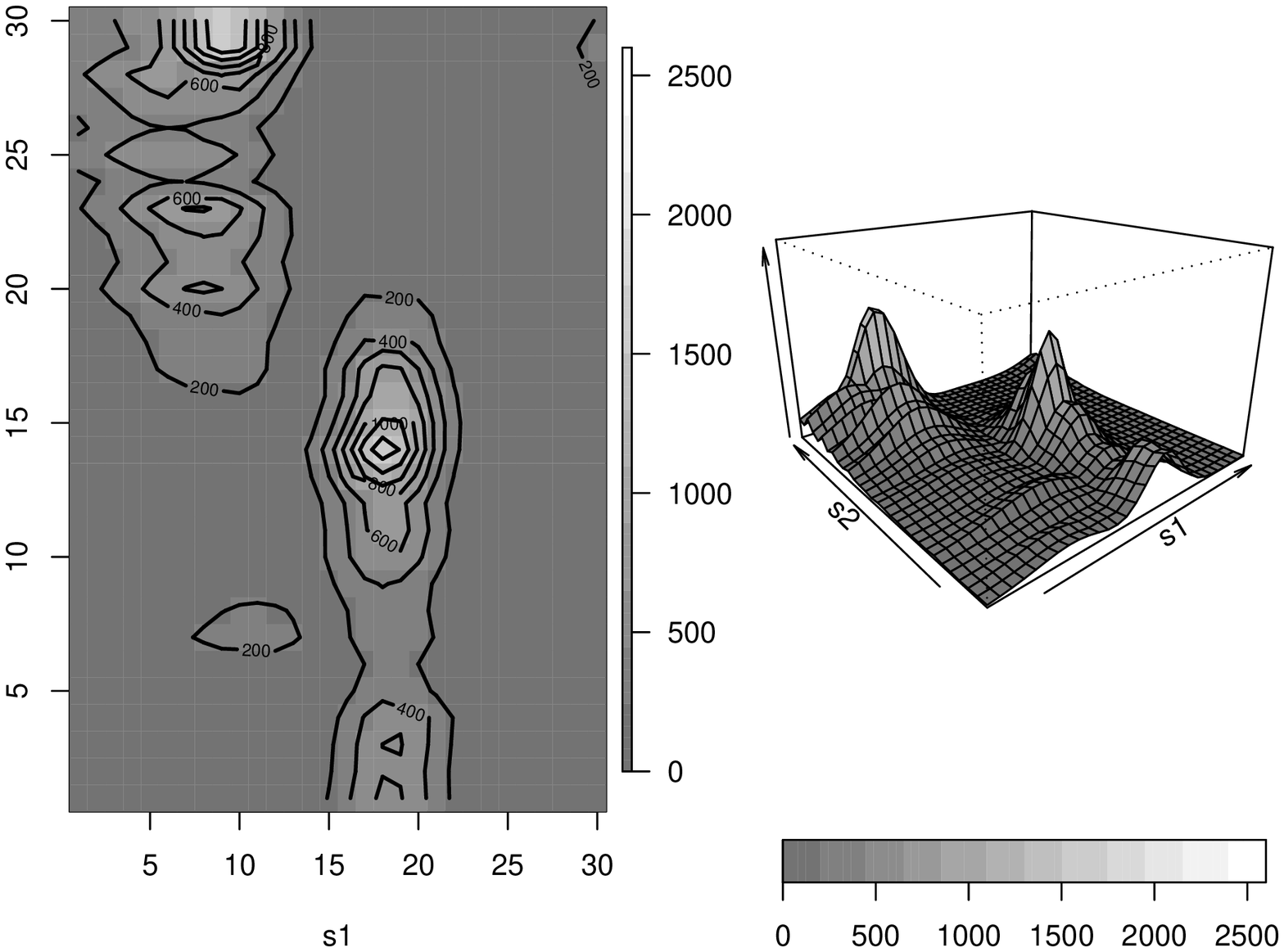}
\vspace*{-0.8cm}

\includegraphics[scale=0.36]{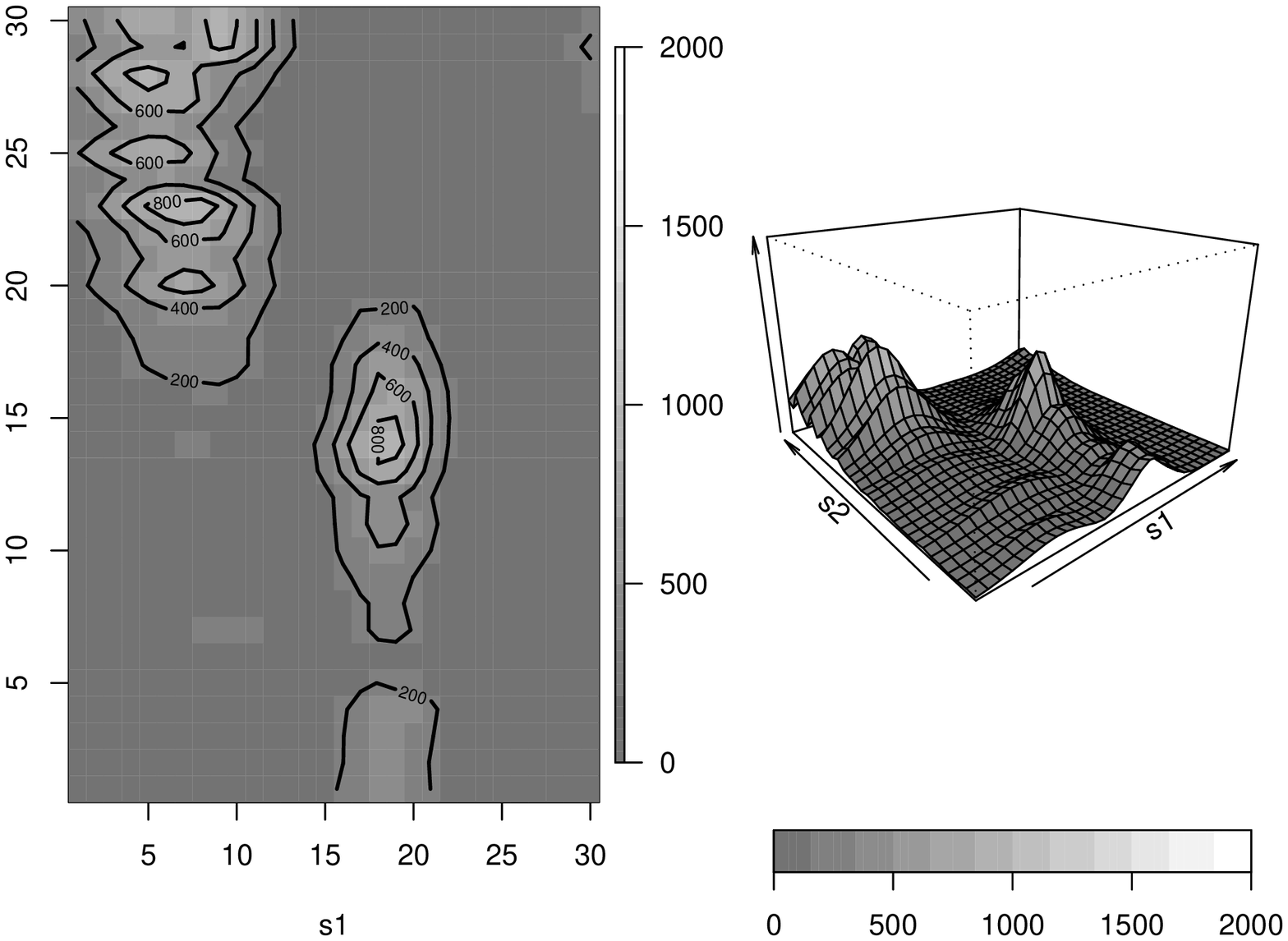}
\vspace*{-0.8cm}

\includegraphics[scale=0.36]{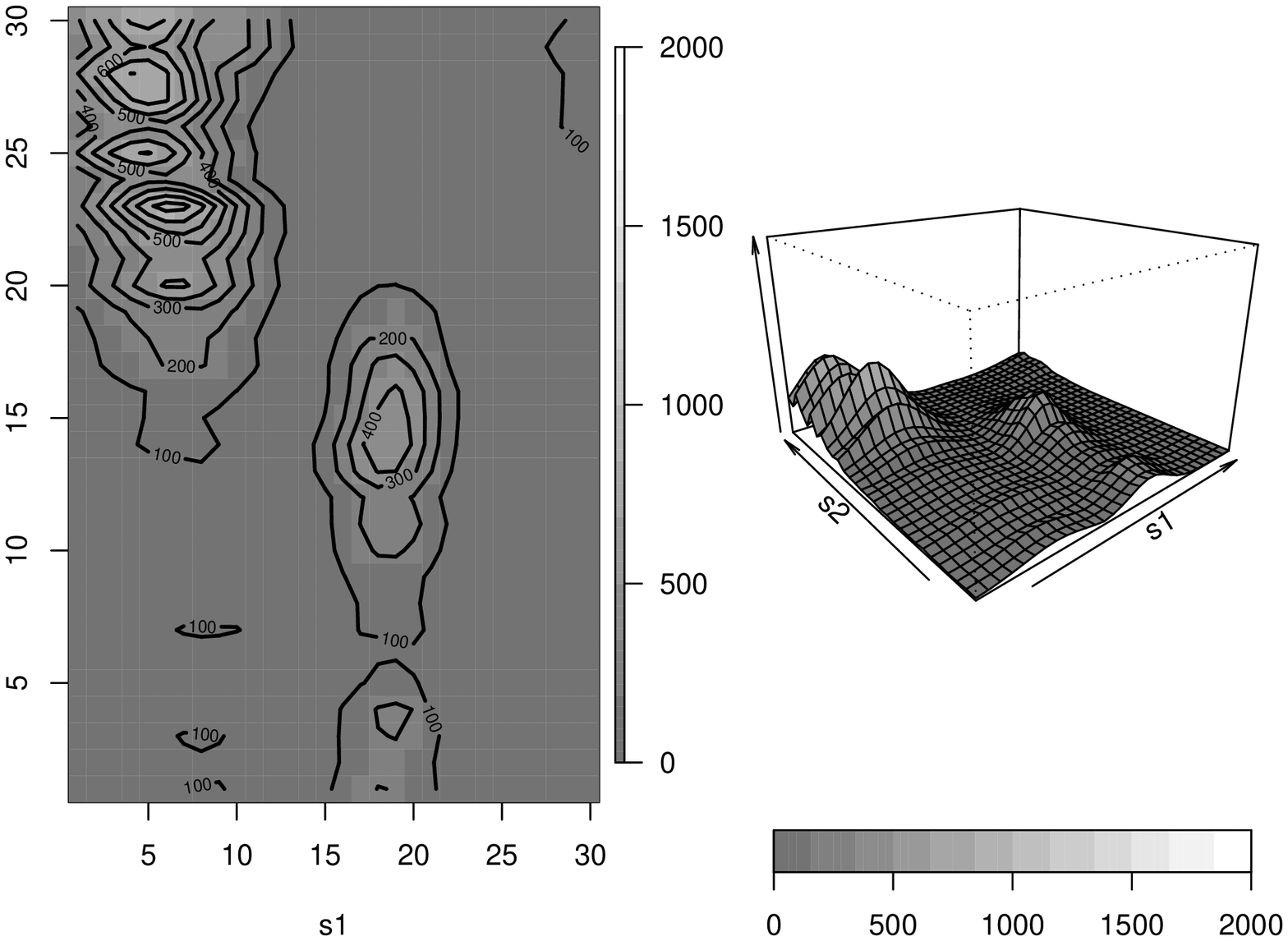}
\captionof{figure}{Simulated anisotropic max-stable random fields with Fr\'echet margins from Example \eqref{ExGneiting} (a=0.03, b=0.03, $\nu=-3/2$, $\gamma=1$) with anisotropic parameters $a_{\text{min}} = 1$, $a_{\text{max}} =3$ and $\alpha=45^{\circ}$.}
\label{anisofields}
\end{center}


\section{Conlusion}
The main objective of this paper was to extend concepts of max-stable random fields in space to the space-time domain. We extended 
the idea of constructing max-stable random fields as pointwise limits of rescaled and transformed Gaussian random fields. (see Kabluchko et al.~\cite{Schlather2})
In a second step, we extended Smith's storm profile model \cite{Smith} and calculated the resulting bivariate distribution functions. 

We showed that the limit assumption on the correlation function in the underlying Gaussian random field relates to the tail dependence coefficient.  
Several examples of spatio-temporal correlation functions and their connection to the tail dependence coefficient have been presented. We extended an assumption on the correlation model known from the analysis of extremes of stationary Gaussian processes and
showed how Gneiting's class of covariance functions \cite{Gneiting} fits in this context. 
Visualizations of our results were shown in form of contour plots of the underlying correlation functions and the corresponding tail dependence coefficients. In addition, we simulated max-stable random fields in space and time using different marginal distributions and an anisotropic correlation function. In particular, geometric anisotropy in the underlying correlation function lead to directional movements in Smith's storm profile model.  

In a forthcoming paper, composite likelihood and other estimation methods for max-stable spatio-temporal random fields are considered.

\noindent
\textbf{Acknowledgment}\\
The first author gratefully acknowledges the support by the International Graduate School of Science and Engineering (IGSSE) of the Technische Universit\"at M\"unchen.

\bibliographystyle{plain}
\bibliography{bibtex_spacetime}

\begin{appendix}
\section{Derivation of the bivariate distribution function for the space-time Smith model}\label{Appendix1}
\begin{proof}{(Theorem \ref{bivariateSmith})}
Since space and time are independent, we can write $$f_0(\bs{z},x) = f_1(\bs{z})f_2(x),\quad \bs{z}\in\bbr^2,\ x\in \bbr $$
where $f_1$ is the density of a bivariate normal distribution with mean $\bs{0}$ and covariance matrix $\Sigma$,
and $f_2$ is the denstiy of a normal distribution with mean $0$ and variance $\sigma_3^2$.
Starting from equation \eqref{smithK} with $K=2$, we obtain
\begin{align*}
F(y_1,y_2) ={} & \exp\left\{-\int\limits_{-\infty}^{\infty}\int\limits_{-\infty}^{\infty}\int\limits_{-\infty}^{\infty}\left(\frac{f_0(\bs{z},x)}{y_1}\right) \vee \left(\frac{f_0(\bs{z}-\bs{h},x-u)}{y_2}\right)d\bs{z}dx\right\}\\
={} & \exp\left\{-\int\limits_{-\infty}^{\infty}\int\limits_{-\infty}^{\infty}\int\limits_{-\infty}^{\infty}\frac{f_0(\bs{z},x)}{y_1}\mathds{1}\left\{\frac{f_0(\bs{z},x)}{y_1} \geq \frac{f_0(\bs{z}-\bs{h},x-u)}{y_2}\right\}d\bs{z}dx\right. \\
&-\left. \int\limits_{-\infty}^{\infty}\int\limits_{-\infty}^{\infty}\int\limits_{-\infty}^{\infty}\frac{f_0(\bs{z}-\bs{h},x-u)}{y_2}\mathds{1}\left\{\frac{f_0(\bs{z}-\bs{h},x-u)}{y_2} \geq \frac{f_0(\bs{z},x)}{y_1}\right\}d\bs{z}dx\right\} \\
={} & \exp\left\{-\text{(I)}-\text{(II)}\right\}
\end{align*}
\begin{align*} 
\text{(I)} 
&=\int\limits_{-\infty}^{\infty}f_2(x)\int\limits_{-\infty}^{\infty}\int\limits_{-\infty}^{\infty}\frac{f_1(\bs{z})}{y_1}\mathds{1}\left\{\frac{f_1(\bs{z})f_2(x)}{y_1} \geq \frac{f_1(\bs{z}-\bs{h})f_2(x-u)}{y_2}\right\}d\bs{z}dx \\
&= \int\limits_{-\infty}^{\infty}f_2(x)\frac{1}{y_1}\mathbb{E}\left[\mathds{1}\left\{\frac{f_1(\bs{Z})f_2(x)}{y_1} \geq \frac{f_1(\bs{Z}-\bs{h})f_2(x-u)}{y_2}\right\}\right]dx, 
\end{align*}
where $\bs{Z}$ has a normal density with mean $\bs{0}$ and variance $\Sigma$.
Now note that 
\begin{align*}
& \frac{f_1(\bs{Z})f_2(x)}{y_1} \geq \frac{f_1(\bs{Z}-\bs{h})f_2(x-u)}{y_2} 
\Leftrightarrow f_1(\bs{Z}) \geq f_1(\bs{Z}-\bs{h})\frac{y_1}{y_2}\frac{f_2(x-u)}{f_2(x)} \\
&\Leftrightarrow (2\pi)^{-d/2}\left|\Sigma\right|^{-1}\exp\left\{-\frac{1}{2}\bs{Z}^{T}\Sigma^{-1}\bs{Z}\right\}
\geq (2\pi)^{-d/2}\left|\Sigma\right|^{-1}\exp\left\{-\frac{1}{2}(\bs{Z}-\bs{h})^{T}\Sigma^{-1}(\bs{Z}-\bs{h})\right\}\frac{y_1}{y_2}\frac{f_2(x-u)}{f_2(x)}  \\
&\Leftrightarrow \bs{Z}^{T}\Sigma^{-1}\bs{Z} < \bs{Z}^{T}\Sigma^{-1}\bs{Z} - 2\bs{Z}^{T}\Sigma^{-1}\bs{h} + \bs{h}^{T}\Sigma^{-1}\bs{h} -2\log\left(\frac{y_1}{y_2}\right) - 2\log\left(\frac{f_2(x-u)}{f_2(x)}\right) \\
&\Leftrightarrow \bs{Z}^{T}\Sigma^{-1}\bs{h} \leq \frac{1}{2}\bs{h}^{T}\Sigma^{-1}\bs{h} - \log\left(\frac{y_1}{y_2}\right) - \log\left(\frac{f_2(x-u)}{f_2(x)}\right). 
\end{align*}
The random variable $\bs{Z}^{T}\Sigma^{-1}\bs{h} =: \bs{Z}^{T}B$ is normally distributed with mean $\bs{0}$ and variance 
$$B^{T}\Sigma B = \bs{h}^{T} \Sigma^{-1}\Sigma\Sigma^{-1}\bs{h} = \bs{h}^{T}\Sigma^{-1}\bs{h}.$$
Since $f_2$ is the density of a zero mean normal distribution with variance $\si_3^2$, we obtain
\begin{align*}
\log\left(\frac{f_2(x-u)}{f_2(x)}\right) &= -\frac{1}{2\sigma_3^2}(u^2-2ux). \\
\end{align*}
With $a(\bs{h}) = (\bs{h}^{T}\Sigma^{-1}\bs{h})^{1/2}$ and $\bs{Z}^T\Sigma^{-1}\bs{h}/a(\bs{h}) \sim \mathcal{N}(0,1)$ it follows
\begin{align*}
& P\left(\bs{Z}^T\Sigma^{-1}\bs{h} \leq \frac{1}{2}\bs{h}^{T}\Sigma^{-1}\bs{h}^{T} - \log\left(\frac{y_1}{y_2}\right) - \log\left(\frac{f_2(x-u)}{f_2(x)}\right) \right) \\
&= P\left(\frac{\bs{Z}^T\Sigma^{-1}\bs{h}}{a(\bs{h})} \leq \frac{a(\bs{h})}{2} - \frac{\log(y_1/y_2)}{a(\bs{h})} + \frac{1}{2\sigma_3^2 a(\bs{h})}(u^2-2ux)\right)\\
&= \Phi\left(\frac{a(\bs{h})}{2} + \frac{\log(y_2/y_1)}{a(\bs{h})} + \frac{u^2-2ux}{2\sigma_3^2 a(\bs{h})}\right).
\end{align*} 
Altogether, for independent random variables $N$ and $X$ with $N$ standard normally distributed and $X$ normally distributed with mean $0$ and variance $\sigma_3^2$, it holds
\begin{align*}
\text{(I)} 
&= \frac{1}{y_1}\int\limits_{-\infty}^{\infty} f_2(x)\Phi\left(\frac{a(\bs{h})}{2} + \frac{\log(y_2/y_1)}{a(\bs{h})} + \frac{u^2}{2\sigma_3^2a(\bs{h})}- \frac{u}{\sigma_3^2 a(\bs{h})} x\right)dx \\
&= \frac{1}{y_1}P\left(N + \frac{u}{\sigma_3 a(\bs{h})} \frac{X}{\sigma_3} \leq \frac{a(\bs{h})}{2}+ \frac{\log(y_2/y_1)}{a(\bs{h})} + \frac{u^2}{2\sigma_3^2a(\bs{h})} \right) \\
&= \frac{1}{y_1}\Phi\left(\frac{\frac{a(\bs{h})}{2}+ \frac{\log(y_2/y_1)}{a(\bs{h})} + \frac{u^2}{2\sigma_3^2a(\bs{h})}}{\sqrt{1+\frac{u^2}{\sigma_3^2 a(\bs{h})^2}}}\right) = \frac{1}{y_1}\Phi\left(\frac{2\sigma_3^2\log(y_2/y_1)+\sigma_3^2a(\bs{h})^2+u^2}{2\sigma_3\sqrt{\sigma_3^2a(\bs{h})^2 + u^2}}\right), \\
\end{align*}
since $N + u/(\sigma_3a(\bs{h}))(X/\sigma_3)$ is normally distributed with mean $0$ and variance $1 + u^2/(\sigma_3^2a(\bs{h})^2)$. \\
Analogously to $\text{(I)}$, using the substitution $\bs{Z} \to \bs{Z} + \bs{h}$, we obtain the second term (II) in \eqref{spacetimeds}.
\end{proof}

\end{appendix}

\end{document}